\documentclass[
superscriptaddress,
twocolumn,
amsmath,amssymb,
aps,
pra,
floatfix,
]{revtex4-2}

\usepackage{mathtools}
\usepackage{amsmath}
\usepackage{amssymb}
\usepackage{cases}
\usepackage{float}
\makeatletter
\newcommand{\ssymbol}[1]{^{\@fnsymbol{#1}}}
\makeatother
\usepackage{mleftright} \mleftright 

\usepackage{scalerel,stackengine}

\usepackage{color}

\usepackage{tensor}

\usepackage{braket}

\allowdisplaybreaks[1] 

\usepackage{xifthen} 
\usepackage{xspace} 
\usepackage[section]{placeins} 

\usepackage{siunitx} 
\usepackage{graphicx} 
\usepackage{tikz} 
\usepackage{xcolor} 
\usetikzlibrary{decorations.pathmorphing} 
\usepackage{soul,xcolor}

\usepackage{comment}

\usepackage{url} 
\usepackage[colorlinks=true, linkcolor=blue, citecolor=green]{hyperref}
\hypersetup{unicode=true} 
\hypersetup{hidelinks} 
\hypersetup{bookmarksopen=true, bookmarksopenlevel=2} 
\hypersetup{pdftitle={title}}
\hypersetup{pdfauthor={Tom Schmit}}
\hypersetup{pdfsubject={}}
\hypersetup{pdfkeywords={}}
\hypersetup{pdflang=en} 
\hypersetup{pdfdisplaydoctitle=true} 
\hypersetup{pdfduplex=DuplexFlipLongEdge} 

\newcommand{\ii}{{\rm i}}

\begin{document}
    \title{Master Equation for a Quantum Gas of Polarizable Particles in Cavities}
	\author{Tom Schmit}
    \affiliation{Theoretische Physik, Universität des Saarlandes, D-66123 Saarbr{\"u}cken, Germany}
	\author{Catalin-Mihai Halati}
    \affiliation{Department of Quantum Matter Physics, University of Geneva, Quai Ernest-Ansermet 24, 1211 Geneva, Switzerland}
    \affiliation{Max Planck Institute for the Physics of Complex Systems, N\"othnitzer Str.~38, 01187 Dresden, Germany}
        \author{Tobias Donner}
    \affiliation{Institute for Quantum Electronics, Eidgen{\"o}ssische Technische Hochschule Z{\"u}rich, Otto-Stern-Weg 1, CH-8093 Zurich, Switzerland}     
        \author{Giovanna Morigi}
    \affiliation{Theoretische Physik, Universität des Saarlandes, D-66123 Saarbr{\"u}cken, Germany}
        \author{Simon B. J{\"a}ger}
    \affiliation{Physikalisches Institut, University of Bonn, Nussallee 12, 53115 Bonn, Germany}

\begin{abstract}
Quantum gases of atoms and molecules in optical cavities offer a formidable laboratory for studying the out-of-equilibrium dynamics of open quantum systems with long-range interactions. Long-range interactions are here mediated by multiple scattering of cavity photons and can induce the formation of quantum structures in space and time. Control of these dynamics requires a detailed understanding of all relevant mechanisms at play. Due to the strong correlations induced by light, however, perturbative theoretical models, which reduce the number of degrees of freedom, do not correctly capture the regime where the interplay of photon-mediated long-range forces and quantum fluctuations of light and matter become significant, such as across the transition to self-organization. In this work, we present the derivation of an effective Lindblad master equation for the dynamics of the sole motional variables of polarizable particles, such as atoms or molecules, that dispersively couple to cavity fields. The master equation is valid even for relatively large intracavity photon numbers, and is apt to study both the steady-state regime and the out-of-equilibrium dynamics where quantum fluctuations of the field seed the onset of macroscopic coherences. We validate the theoretical description by showing that it captures the dynamics across a wide temperature interval, from Doppler cooling down to the ultra-cold regime, and from weak to strong cavity-mediated interactions. Our theory provides a powerful framework for the description of cavity-induced dynamics of quantum matter. In doing so, it permits to connect models of statistical mechanics with cavity-QED experimental platforms, thus enabling quantum simulation of long-range-interacting matter. 
\end{abstract}

\date{\today}
\maketitle

\section{Introduction}
\label{sec:Introduction}
Many-body cavity quantum electrodynamics is a suggestive label for the strong-coupled dynamics of systems with many particles and light in confined geometries, such as Fabry-P{\'e}rot cavities and fibers~\cite{Ritsch:2013, Chang:2018, Mivehvar2021Cavity}. This field of research is motivated and pushed forward by advanced experimental platforms working in this regime~\cite{Ritsch:2013, Mivehvar2021Cavity}. One salient feature of these systems is multiple photon scattering, which can transfer energy and imprint long-range correlations between the individual emitters. A further important characteristic is photon losses, which renders the dynamics intrinsically dissipative. In the strong-coupling regime, where the scatterers prevailingly emit into the cavity mode while emitter-photon interactions only couple resonant electronic levels~\cite{Giannelli:2024}, non-trivial phenomena can be observed by means of external drives, realizing exemplary driven-dissipative many-body dynamics. The wealth of observed collective phenomena includes cavity-induced laser cooling~\cite{Schleier-Smith:2011,Wolke:2012}, self-organization in ordered spatial patterns~\cite{Black:2003,Baumann2010Dicke}, lasing induced by collective scattering~\cite{Kruse:2003,Bohnet:2012}, supersolidity~\cite{Leonard:2017,Morales2017Coupling}, and time-crystal dynamics~\cite{Kessler:2021,Kongkhambut:2022}. 

The ability to describe these dynamics in full detail sheds light onto these cooperative phenomena and permits the identification of microscopic control tools of collective behavior, thereby opening novel pathways toward quantum simulation of strongly correlated matter and to quantum computing~\cite{Marsch:2021, Mivehvar2021Cavity,Halati:2025b}. Recently, increasing efforts are being invested into exporting these concepts to molecules and solids~\cite{Delic:2019, Ashida:2020, Kalaivanan:2021, Schlawin:2022, Jarc:2023}, with the perspective of designing novel states of matter and dynamics using photon scattering~\cite{Mivehvar2021Cavity}. However, state-of-the-art studies of molecules in cavities typically consider one up to a few scatterers and discard the center-of-mass motion~\cite{Haugland:2020,Monzel:2024,Schnappinger:2023}, while the theoretical description of the full dynamics becomes quickly challenging, especially when including the mechanical effects of light. For all these reasons, the development of effective models with tested validity is of central importance. 

\begin{figure*}
    \centering
    \includegraphics[width=1\textwidth]{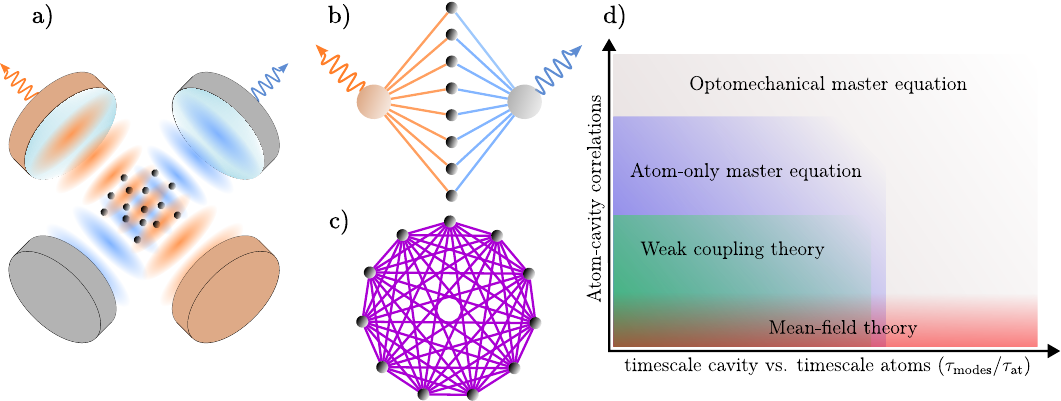}
    \caption{(a) Characteristic setup of many-body CQED that consists of polarizable particles (e.g., atoms or molecules) strongly coupled to one or multiple optical resonators (here two modes are illustrated). (b) In the optomechanical regime, the excited electronic states are only virtually populated: Photon absorption and emission coherently couple the motion of the particles (black circles) to the cavity field modes (large shaded circles), which in turn can dissipate photons via emission through the mirrors (wavy arrows). (c) The resulting atom-only dynamics is illustrated by a globally connected network, where the nodes represent the particles and the links correspond to photon-mediated interactions. (d) Regime of validity for different approaches to describe the optomechanical master equation. Mean-field approaches fail to describe atom-cavity correlations. Weak-coupling theories have limited validity for describing atom-cavity correlations and rely on a timescale separation. The focus of this work is to identify the regimes in which the dynamics can be described by atom-only master equations that can capture strong atom-cavity correlations.}
    \label{fig:1}
\end{figure*}

Among several cavity-assisted phenomena, self-organization of atoms into crystalline structures bound by photons is one of the most intriguing~\cite{Black:2003,Kruse:2003,Baumann2010Dicke,Arnold:2012,Brennecke:2013,Mottl:2012,Ritsch:2013,Landig:2016}. These experiments are realized in Fabry-P{\'e}rot cavities in the regime where the internal structure of the scatterers is captured by their polarizability, and the energy they exchange with radiation is solely mechanical~\cite{Domokos:2003}. An illustrative example is shown in Fig.~\ref{fig:1}(a). Several effective theoretical models have been proposed for describing the resulting dynamics~\cite{Domokos:2003,Maschler:2005,Ritsch:2013,Schuetz:2013,Piazza:2013,Mivehvar2021Cavity}. These models are derived starting from the minimal-coupling Hamiltonian in the electric dipole approximation, where spontaneous emission and cavity decay are described by a Born-Markov master equation, and systematically lead to a master equation in which the cavity field directly couples to the particles' external degrees of freedom~\cite{Domokos:2001,Domokos:2003,Schuetz:2013,Ritsch:2013}, as sketched in Fig.~\ref{fig:1}(b) for the setup of subplot (a). We dub this master equation {\it optomechanical}, extending to the microscopic realm a term typically used in the mesoscopic regime~\cite{Aspelmeyer:2014}.

Despite the notable reduction of the configuration space, the theoretical treatment of these optomechanical models remains challenging in many-body settings. Even for a single-mode cavity, the size of the Hilbert space becomes prohibitively large for moderate intracavity photon numbers. As a result, the effective models available in the literature often become invalid in the most interesting regimes---for instance, beyond the paradigmatic Dicke phase transition~\cite{Baumann2010Dicke,Brennecke:2013}. Therefore, the modeling of these dynamics is often founded on phenomenological, albeit plausible, assumptions~\cite{Baumann2010Dicke,DallaTorre:2013,Zupancic2019P}. In addition, the majority of the studies rely on mean-field approximations, discarding fluctuation effects that are relevant at the self-organization transition, or on the weak-coupling regime, whose validity is limited to the normal, disordered phase. Some of these models are described by master equations, which are derived from the optomechanical master equation by perturbatively eliminating the photonic degrees of freedom~\cite{Nimmrichter:2010,Chiacchio:2018}. This procedure further reduces the configuration space and yields a so-called \textit{atom-only} description, in which the only remaining degrees of freedom are atomic. Figure~\ref{fig:1}(c) shows an abstraction of the resulting model for the setup of subplot (a), where the atoms interact via global forces. Atom-only models were also derived using operator-based approaches for ultracold atoms, under the assumption that the atomic temperature is sufficiently low such that the cavity field adiabatically follows the atomic motion, even for relatively large photon numbers~\cite{Larson:2008,Larson:2008a,Fernandez:2010}. Models based on this approach~\cite{Dogra:2016,Niederle:2016,Flottat:2017,Himbert:2019} have been qualitatively reproducing experimental measurements~\cite{Landig:2016}. Subplot (d) illustrates the expected regimes of validity of the various approaches. Recently, the validity of atom-only models has been questioned in several studies~\cite{Damanet:2019,Halati:2020a,Palacino:2021,Orso:2025}, some of which provide numerical evidence of discrepancies between their predictions and the numerical simulations of the corresponding optomechanical master equations. Solving this debate would enable to develop an overarching atom-only framework that permits to effectively describe a wide parameter regime, including the transition to self-organization and the onset of macroscopic, light-induced quantum coherences and entanglement.

In this work, we resolve this debate by rigorously deriving an effective atom-only Lindblad master equation governing the external degrees of freedom of polarizable particles. This is obtained by significantly extending a procedure~\cite{Jaeger:2022} which was validated for spin systems and in the weak-coupling regime~\cite{Jaeger:2022,Jaeger:2023,Jaeger:2024}. The resulting atom-only master equation describes the dynamics of continuous variables, namely, of the motion of an atomic gas induced by the coupling with cavity fields. It accurately captures the relevant atom-field correlations---both in the dispersive~\cite{Chiriaco:2022, Bacciconi:2023} and in the dissipative regime~\cite{Wall:2016, Halati:2020a, Halati:2022, Halati:2025a, Halati:2025b}. It provably remains valid when the cavity-mediated potentials vary across the self-organization transition. Moreover, it provides an overarching and unifying framework, as it correctly reproduces established models that are either valid in the limit of vanishing photon numbers~\cite{Piazza:2014,Chiacchio:2018} or in the regime of large intracavity fields~\cite{Domokos:2001,Mivehvar2021Cavity}, as well as models previously derived through semiclassical treatments~\cite{Schuetz:2013} or operator-based approaches~\cite{Maschler:2005,Larson:2008,Fernandez:2010,Habibian:2013a}. Our master equation thus establishes a rigorous basis for the systematic analysis of stationary states and critical phenomena in many-body cavity-QED systems, as well as their out-of-equilibrium dynamics. As such, it serves as a tool for elucidating the interplay between dissipative and dispersive long-range forces and for identifying the key control parameters that enable tailoring the quantum dynamics of atomic and molecular gases in experiments. Our model is generally applicable to a wide range of experimental setups, including multi-mode cavities, that study patterns induced by strong cavity-mediated interactions~\cite{Baumann2010Dicke,Landig:2016,Leonard:2017,Guo:2021,Wu:2023,Helson:2023,Kroeze:2025,Ho:2025}. For all these reasons, it provides an essential framework for testing and implementing quantum simulators with cavity-QED platforms.

This paper is organized as follows. In Sec.\ \ref{sec:General model}, we first introduce the generic model of atoms or molecules dispersively coupled to modes of a high-finesse cavity. In Sec.\ \ref{Sec:AtomOnly}, we present the derivation of the atom-only master equation. We then determine its specific form and examine its predictive capabilities in the limit of vanishing intracavity photon numbers (Sec.\ \ref{Sec:weak}) as well as when intracavity fields increase across the self-organization phase transition (Sec.\ \ref{Sec:strong}), using experimentally relevant dynamics as case studies. The conclusions are drawn in Sec.\ \ref{Sec:Conclusions}, while the appendices provide details of the derivations in Secs.\ \ref{Sec:AtomOnly}, \ref{Sec:weak}, and \ref{Sec:strong}.

\section{Optomechanical master equation for many-body CQED}
\label{sec:General model}

In this section, we review the assumptions at the basis of the optomechanical master equation, describing the dynamics of polarizable particles strongly coupled to optical fields via the mechanical effects of light. The purpose is to introduce the individual parameters and processes determining the dynamics. These impose important constraints that shall be accounted for in Sec.\ \ref{Sec:AtomOnly} by deriving the atom-only master equation. Readers who are familiar with the formalism may choose to skip this section.

\subsection{The Model}

We consider $N$ identical polarizable particles of mass $m$ interacting with $M$ optical modes in a confined volume. An exemplary situation is shown in Fig.\ \ref{fig:1}(a), depicting a gas interacting with $M=2$ cavity modes. The particles can be atoms or molecules, possessing an optical dipole transition that couples to the optical modes. The state of the system, composed of the particles' internal and external degrees of freedom and the optical modes, is described by the density operator $\hat{\rho}$. In quantum optical settings, where dissipation manifests as spontaneous emission and photon losses at the mirrors or in fibers, the equation of motion for $\hat{\rho}$ takes the form of a Born-Markov master equation:
\begin{equation}
    \frac{\partial}{\partial t}\hat{\rho} = \frac{1}{{\rm i}\hbar}[\hat{H},\hat{\rho}] +\mathcal{L}_{\rm diss}\hat{\rho}\label{eq:full_master_equation}\,.
\end{equation}
Here, $\mathcal{L}_{\rm diss}$ is the superoperator describing the incoherent dynamics, and the Hamiltonian $\hat{H}$ governs the coherent coupled dynamics in the electric dipole and rotating-wave approximation. Below, we introduce the individual terms and discuss the relevant timescales associated with each process.

\subsection{Atomic Dynamics}

We first focus on the atomic dynamics, and introduce later the coupling with the cavity modes. The particles' external degrees of freedom are the canonically-conjugated position and momentum, $\hat{\vec{r}}_j$ and $\hat{\vec{p}}_j$ ($j=1,\ldots,N$). In what follows, we assume that the dipolar transition is composed of two levels, namely, the ground state $\ket{g}$ and the excited state $\ket{e}$. We remark that the two-level approximation is here used for convenience but is not essential, as our focus lies in the regime of coherent scattering, where the internal structure is captured by the polarizability. 

The atomic Hamiltonian is the sum of the energy for the internal and external degrees of freedom, and includes the interaction with a transverse pump, driving the particles: $\hat H_{\rm at}=\hat{H}_\mathrm{internal}+\hat{H}_\mathrm{ext}+\hat{H}_\mathrm{pump}$. The first two terms read  
\begin{align}
    \hat{H}_\mathrm{internal} &= \sum_{j=1}^N\hbar(-\Delta_\mathrm{a}) \hat{\sigma}_j^\dagger\hat{\sigma}_j\,,\\
    \hat{H}_\mathrm{ext} &= \sum_{j=1}^N\left(\frac{\hat{\vec{p}}_j^2}{2m} + W(\hat{\vec{r}}_j)\right)\label{eq:def_H_ext}\,,
\end{align}
where $\hat{\sigma}_j=|g\rangle_j\langle e|$ and $\hat{\sigma}_j^\dagger$ is its adjoint, $W(\hat{\vec{r}})$ is a generic conservative potential confining the atomic center-of-mass motion, and $\Delta_\mathrm{a} = \omega_\mathrm{p} - \omega_\mathrm{a}$ is the detuning between the frequency $\omega_\mathrm{p}$ of the transverse pump and the atomic transition frequency $\omega_\mathrm{a}$. The transverse pump is assumed to be a classical monochromatic field whose interaction with the atomic dipole is described by the Hamiltonian within the electric dipole approximation:
\begin{equation}
     \hat{H}_\mathrm{pump}=\sum_{j=1}^N\hbar\,\Omega f_\mathrm{p}(\hat{\vec{r}}_j)\hat \sigma_j^\dagger +{\rm h.c.}\,,
\end{equation}
where ${\rm h.c.}$ stands for hermitian conjugate, $\Omega$ is a constant Rabi frequency, and $f_\mathrm{p}(\hat{\vec{r}})$ is the dimensionless spatial profile of the pump field at the atomic position, with $\Vert f_\mathrm{p}(\hat{\vec{r}})\Vert\leq1$.

Spontaneous emission of the atomic dipole into the modes external to the resonators is described by the Lindblad superoperator~\cite{Dalibard:1985,Marte:1993}
\begin{equation}
    \mathcal L_{\gamma}\hat{\rho}=\gamma \sum_{j=1}^N\left(2\hat \sigma_j\hat{\bar \rho}_j\hat \sigma_j^\dagger-[\hat \sigma_j^\dagger\hat \sigma_j,\hat{\rho}]_+\right)\,.
\end{equation}
Here, $[\hat{A},\hat{B}]_+ = \hat{A}\hat{B} + \hat{B}\hat{A}$ is the anti-commutator and $\hat{\bar \rho}_j$ denotes the density operator that includes the mechanical effects of the spontaneously emitted photon on particle $j$~\cite{meystre:2007}:
\begin{equation}
    \label{eq:radiative_losses_superoperator}
    \hat{\bar \rho}_j = \int_{V_\perp}{\rm d}^3k\, \mathcal{N}(\vec k){\rm e}^{-\ii\vec{k}\cdot\hat{\vec{r}}_j}\,\hat{\rho}\,{\rm e}^{\ii\vec{k}\cdot\hat{\vec{r}}_j}\,.  
\end{equation}
The scalar function $\mathcal{N}(\vec k)$ is the probability density for emitting a photon with wave vector $\vec k$, normalized as $\int{\rm d}^3k\,\mathcal{N}(\vec k)=1$~\cite{Lethokov:1981}. The integral runs over all wave vectors $\vec{k} \in V_\perp$, corresponding to emission into the solid angle outside of the resonators~\cite{Vitali:2006}. For simplicity, we neglect collective decay processes into the modes external to the cavities, assuming that the average interatomic distance is much larger than the corresponding wavelengths.

\subsection{Cavity Dynamics}

The atomic transition couples to $M$ cavity modes with frequencies $\omega_n$, forming a discrete spectrum ($n=1,\ldots, M$). We denote by $\hat{a}_n$ and $\hat{a}_n^\dagger$ the annihilation and creation operators of a quantum of energy $\hbar\omega_n$, with $[\hat{a}_n,\hat{a}_{n'}^\dagger]=\delta_{n,n'}$. Their free evolution is governed by the Hamiltonian (in the frame rotating at the transverse-pump frequency)
\begin{equation}
    \hat{H}_\mathrm{modes} = \sum_{n=1}^M\hbar (-\Delta_n) \hat{a}_n^\dagger \hat{a}_n\,,
\end{equation}
with $\Delta_n = \omega_\mathrm{p} - \omega_n$. For the sake of generality, we also consider the Hamiltonian term describing a longitudinal pump with frequency $\omega_\mathrm{p}'$ of the cavity modes:
\begin{equation}
    \hat{H}_{\rm pump}^\mathrm{(c)}= \sum_{n=1}^M\hbar\,\zeta_n^*{\rm e}^{-{\rm i}\delta\omega_\mathrm{p}t} \hat a_n+{\rm h.c.}\,,
\end{equation}
where $\zeta_n$ is the strength of the pump on mode $n$ and has the dimensions of a frequency, and $\delta\omega_\mathrm{p}=\omega_\mathrm{p}-\omega_\mathrm{p}'$ is the frequency of the longitudinal pump in the reference frame rotating with the transverse-pump frequency $\omega_\mathrm{p}$.

In what follows, we also assume that the incoherent dynamics of the cavity modes is solely due to photon losses at the cavity mirrors:
\begin{equation}
    \label{eq:L_kappa}
    \mathcal{L}_\kappa \hat{\rho} = \sum_{n=1}^M\kappa_n\mathcal{D}[\hat{a}_n]\hat{\rho}\,,
\end{equation}
with $\kappa_n$ the decay rate of mode $n$ and
\begin{equation}
    \label{eq:Dissipator}
    \mathcal{D}[\hat{O}]\hat{\rho}=2\hat{O} \hat{\rho}\hat{O}^\dagger - [\hat{O}^\dagger \hat{O},\hat{\rho} ]_+\,.    
\end{equation}
Note that we neglect the incoherent absorption of photons, which is justified for optical frequencies at room temperature. Processes involving photon absorption by the mirrors are also not included, although they can often be included by straightforwardly extending this formalism, as shown in Ref.\ \cite{Giannelli:2018}. 

\subsection{Atom-Cavity Coupling}

The interaction between cavity fields and atoms is a generalization of the Tavis-Cummings Hamiltonian to the multi-mode case~\cite{Larson:2024}:
\begin{equation}
    \hat{V}_\mathrm{int} = \sum_{n=1}^M\sum_{j=1}^N \hbar\,g_n(\hat{\vec{r}}_j)\hat{\sigma}_j^\dagger\hat{a}_n +{\rm h.c.}\,,
    \label{Eq:V:0}
\end{equation}
which includes the spatial dependence of the atom-photon coupling, $g_n(\hat{\vec{r}}) = g_{n}f_n(\hat{\vec{r}})$. Here, $g_n$ is the vacuum Rabi frequency of cavity mode $n$, and $\Vert f_{n}(\hat{\vec{r}})\Vert \leq 1$ is the spatial mode function.

\subsection{Optomechanical Master Equation}
\label{Sec:Optomechanical_master_equation}

The optomechanical master equation is derived in the regime where the atoms' (or molecules') excited states can be eliminated from the dynamical equation of the atomic center-of-mass motion and the cavity fields, see, for instance, Refs.\ \cite{Mivehvar2021Cavity,Domokos:2001,Schuetz:2013}. This regime allows to replace the particles with their polarizability and requires that the rate $|\Delta_\mathrm{a}+{\rm i}\gamma|$ exceeds the coupling strengths, $|g_n|\sqrt{n_n},|\Omega|\ll |\Delta_\mathrm{a}+{\rm i}\gamma|$, with $n_n=\langle \hat{a}_n^\dagger \hat{a}_n\rangle$ the average photon number in mode $n$~\cite{Larson:2008,Schuetz:2013,Keller:2018}. In order to get a set of equations where the cavity and external variables are directly coupled, the internal degrees shall evolve on a much faster timescale. For the cavity, this requires $|\Delta_n + \ii \kappa_n|,|\delta\omega_\mathrm{p}|\ll |\Delta_\mathrm{a}+{\rm i}\gamma|$. For the external motion, the slow rate is determined by the mean motional energy and by the recoil energy at the wavelength of the light, as we will extensively discuss in the next section. 

The resulting master equation governs the dynamics of the density operator $\hat{\varrho}$ defined in the Hilbert space of the cavity fields and particle motion~\cite{Schuetz:2013}:
\begin{equation}
    \frac{\partial}{\partial t}\hat{\varrho} = \frac{1}{\ii \hbar}[\hat H_{\rm eff}, \hat{\varrho}] + \mathcal{L}_\kappa \hat{\varrho} + \mathcal{L}_{\gamma,\mathrm{eff}}\hat{\varrho}\label{eq:opto_master_equation}\,,
\end{equation}
where $\hat H_{\rm eff}=\hat{H}_\mathrm{ext} + \hat{H}_\mathrm{modes} + \hat H_{\rm pump}^{(c)}+\hat{V}_\mathrm{eff}$, and the optomechanical interaction between photonic fields and external degrees of freedom reads
\begin{equation}
\label{eq:V:int}
    \hat{V}_\mathrm{eff} = \hbar \frac{\Delta_\mathrm{a}^2}{\Delta_\mathrm{a}^2 + \gamma^2} \sum_{j=1}^N\frac{1}{\Delta_\mathrm{a}} \hat{F}_j^\dagger \hat{F}_j\,.
\end{equation}
Here,
\begin{equation}
    \label{Eq:F}
    \hat{F}_{j} = \Omega f_\mathrm{p}(\hat{\vec r}_j) + \sum_{n=1}^Mg_n f_n(\hat{\vec r}_j) \hat{a}_n
\end{equation}
is the coherent sum of the transverse pump and cavity fields. The Hamiltonian term in Eq.~\eqref{eq:V:int} is an effective interaction between the cavity modes that depends on the atomic spatial distribution. At the same time, it is also an effective mechanical potential whose depth is determined by the sum of the classical pump and the quantum cavity fields. The form of Eq.\ \eqref{Eq:F} also shows that the individual atomic transitions act as beam splitters, coherently mixing pump and cavity fields, and thereby enabling interference between them~\cite{Zippilli_2004}. As a result, for certain atomic distributions, the potential \eqref{eq:V:int} can vanish, see Refs.\ \cite{Fernandez:2007, Habibian:2013b, Reimann2015Cavity, Zupancic2019P, Baumgaertner:2025}. 

We rewrite $\hat{V}_\mathrm{eff}=\Delta_\mathrm{a}^2/(\Delta_\mathrm{a}^2+\gamma^2)\,(\hat V_\mathrm{p}+\hat V_{\rm cav})$ as the sum of the mechanical potential due to coherent scattering of pump photons, $\hat V_\mathrm{p}=\sum_j\hbar\,(|\Omega|^2/\Delta_\mathrm{a})(f_\mathrm{p}(\hat{\vec{r}}_j))^\dagger f_\mathrm{p}(\hat{\vec{r}}_j)$, and of all other coherent scattering processes involving the cavity fields,
\begin{equation}
 \hat V_{\rm cav} = \sum_{n,m=1}^M\hbar\,U_{nm} \hat{a}_n^\dagger \hat{\Theta}_{nm}\hat{a}_m + \sum_{n=1}^M \hbar\,\left(\eta_n^*\hat{\Theta}_{n {\rm p}}^\dagger\hat{a}_n + \mathrm{h.c.}\right)\label{eq:H_eff_explicit}\,.  
\end{equation}
For compactness, we have introduced $\eta_n= \Omega g_n^*/\Delta_\mathrm{a}$, scaling the scattering processes which mix pump and cavity photons, while $U_{nm} = g_{n}^*g_{m}/\Delta_\mathrm{a}$ is the amplitude of processes where a photon is scattered from cavity mode $m$ to cavity mode $n$ (``cross-talking"). The operators $\hat{\Theta}$ depend on the positions of the particles through the mode functions of the fields:
\begin{equation}
    \label{eq:def_Theta_operator}
    \begin{aligned}
        \hat{\Theta}_{n {\rm p}} &= \sum_{j=1}^N f_n^\dagger(\hat{\vec r}_j)f_\mathrm{p}(\hat{\vec r}_j)\,,\\
        \hat{\Theta}_{nm} &= \sum_{j=1}^N f_n^\dagger(\hat{\vec r}_j)f_m(\hat{\vec r}_j)\,,
    \end{aligned}
\end{equation}
with $n,m=1,\dots,M$. Their specific form determines the density patterns that maximize the corresponding scattering processes. This form is valid also when each cavity mode is pumped by means of a multi-mode laser, each driving a different dipolar transition~\cite{Torggler:2017,Keller:2017}.  In this case, the cross-talking terms $\hat{\Theta}_{nm}$ vanish.

Within this description, spontaneous atomic decay now gives rise to an effective incoherent dynamics of the cavity photons and atomic motion, described by the superoperator~\cite{Vitali:2006, Schuetz:2013}
\begin{eqnarray}
    \mathcal{L}_{\gamma,\mathrm{eff}}\hat{\varrho} =\frac{\gamma}{\Delta_\mathrm{a}^2 + \gamma^2} \sum_{j=1}^N
    \left(2\hat{F}_j\hat{\bar\varrho}_j\hat{F}_j^\dagger  - [\hat{F}_j^\dagger\hat{F}_j,\hat{\varrho}]_+\right)\nonumber\,,
\end{eqnarray}
with $\hat{\bar\varrho}_j$ defined similarly to Eq.\ \eqref{eq:radiative_losses_superoperator}. Interestingly, incoherent scattering can also be enhanced or suppressed by interference between the pump and cavity fields, see Ref.\ \cite{Vitali:2006}. In the following, we will discard the incoherent dynamics due to spontaneous emission by assuming $\gamma\ll |\Delta_\mathrm{a}|$, typically referred to as \textit{dispersive regime}.

The optomechanical master equation, Eq.\ \eqref{eq:opto_master_equation}, is the model at the basis of most studies of cavity self-organization and cavity cooling. Despite the elimination of the internal degrees of freedom, the analysis of the dynamics often requires further assumptions. Figure \ref{fig:1}(d) summarizes the various approaches used in the literature and the expected range of validity. A widely adopted approach consists of performing a mean-field approximation of the cavity fields amplitudes, where quantum fluctuations are neglected~\cite{Domokos:2001,Domokos:2003,Kongkhambut_2024}. This approximation is expected to hold for sufficiently large cavity fields, away from bistable points, and can be extended to include corrections due to cavity field fluctuations~\cite{Cormick:2012,Bezvershenko:2021,Tolle:2025}. However, it has limited predictive power in the regime where fluctuations dominate and atom-cavity correlations are important, such as at the phase transition where the particles self-organize in ordered Bragg gratings. 

The atom-only approach simplifies the optomechanical master equation without performing a mean-field approximation. Such a description relies on a timescale separation between particle and cavity dynamics, but it does not assume vanishing correlations between them, as we argue in the following section.

\section{Atom-Only Master Equation}
\label{Sec:AtomOnly}

The atom-only master equation for the dynamics of a gas of polarizable particles is obtained by eliminating the photonic degrees of freedom from the optomechanical master equation. It notably reduces the configuration space to the Hilbert space of the particles' external degrees of freedom and allows us to efficiently describe cavity-mediated interactions and dissipation. For few cavity modes, the cavity mediates global interactions, as illustrated in Fig.~\ref{fig:1}(c). In its general form, the equation is non-local in time and does not simplify the theoretical analysis. Here, we determine the regime in which it can be reduced to a time-local master equation of Lindblad form. We will argue that the resulting Lindblad master equation constitutes a valid quantum model within the regime of a timescale separation between cavity and particle degrees of freedom. We further show that the regime of validity can be systematically extended to the regime of strong interactions and relatively large intracavity photon numbers.

Our derivation extends the procedure of Ref.\ \cite{Jaeger:2022}, which was derived in the weak-coupling regime for spin models. Transferring this method to the continuum of the external motion is nontrivial, as we will discuss. We further highlight that we also extend and verify the approach for strong cavity-mediated interactions. In the following, we begin with the optomechanical master equation \eqref{eq:opto_master_equation} and outline the derivation of the atom-only master equation. Additional details are provided in Appendix~\ref{App:A}. 

\subsection{Displaced Reference Frame}

The optomechanical master equation \eqref{eq:opto_master_equation} in the dispersive regime ($|\Delta_\mathrm{a}| \gg \gamma$) takes the form
\begin{equation}
	\frac{\partial}{\partial t}\hat{\varrho} = \frac{1}{\ii \hbar}[\hat H_{\rm eff}, \hat{\varrho}] + \mathcal{L}_\kappa \hat{\varrho} \label{eq:opto_master_equation_disp}\,.
\end{equation}
The Hamiltonian reads
\begin{equation}
    \label{eq:def_H_eff}
    \hat{H}_{\mathrm{eff}} = \hat{H}_{\rm S}+\hat{H}_{\rm cav}\,,
\end{equation}
where $\hat{H}_{\rm S}$ is the Hamiltonian of the external degrees of freedom in the absence of the coupling with the cavity modes:
\begin{equation}
    \hat{H}_{\rm S} = \sum_{j=1}^N\left(\frac{\hat{\vec{p}}_j^2}{2m}+W(\hat{\vec{r}}_j)+\hbar \frac{|\Omega|^2}{\Delta_\mathrm{a}}f_\mathrm{p}(\hat{\vec{r}}_j)^\dagger f_\mathrm{p}(\hat{\vec{r}}_j)\right)\,, 
\end{equation}
while $\hat{H}_{\rm cav}$ contains the optomechanical coupling between cavity modes and external degrees of freedom:
\begin{equation}
    \hat{H}_{\rm cav}=\sum_{n,m=1}^{M}\hbar\,\hat{a}_n^\dagger\hat{\Omega}_{nm}\hat{a}_m+\sum_{n=1}^{M}\hbar\left(\hat{G}_n^\dagger(t)\hat{a}_n+\hat{a}_n^\dagger\hat{G}_n(t)\right)\,.\label{eq:Hcav}
\end{equation}
The operators are
\begin{equation}
    \begin{aligned}
        \hat{\Omega}_{nm} &= -\delta_{n,m}\Delta_n+U_{nm}\hat{\Theta}_{nm}\,,\\
        \hat{G}_n &= \eta_{n}\hat{\Theta}_{n\mathrm{p}}+\zeta_ne^{i\delta\omega_\mathrm{p}t}\,,
    \end{aligned}
\end{equation}
and they depend on the atomic positions within the cavity modes. 

We first discuss an insightful case, that will lead our derivation. Let us, for the moment, consider particles of infinite mass. In this case, the atomic operators in Eq.\ \eqref{eq:Hcav} become scalar functions, and  the Hamiltonian $\hat{H}_{\rm cav}$ can be brought into a quadratic, diagonal form for the cavity field operators via a unitary transformation. In particular, the second term on the RHS of Eq.\ \eqref{eq:Hcav} is a displacement operator of the cavity field modes, with amplitudes that depend on the external longitudinal pump field and the atomic distribution within the cavity modes. The standard procedure in the derivation of quantum optical master equations would then consist in transforming to a displaced reference frame in which the cavity field modes are in the vacuum state~\cite{cohen1992atom,Gardiner:QNoise}. 

Motivated by this observation, we now return to particles of finite mass and postulate the unitary, time-dependent displacement operator 
\begin{equation}
	\label{eq:D}
	\hat{D}(t) = \exp[\hat{r}(t)]\,,
\end{equation}
with $\hat{r}^\dag=-\hat{r}$, for displacing the cavity modes into the vacuum state. Following the intuition gained from the infinite mass limit, we are led to the following ansatz:
\begin{equation}
    \hat{r}(t) = \sum_{n=1}^{M}(\hat{a}_n^\dagger\hat{\alpha}_{n}(t)-\hat{\alpha}_{n}^\dagger(t)\hat{a}_n)\,,\label{eq:r_general}
\end{equation}
where $\hat{\alpha}_n$ are operators acting on the atomic Hilbert space. Physically, they can be interpreted as source fields for the cavity modes that contain information about the atomic state. Next, we determine the equations of motion for the operators $\hat{\alpha}_n$.

\subsection{Derivation of the Atom-Only Master Equation}

We determine the equations for the operators $\hat{\alpha}_n$ starting from the master equation for the density operator $\hat{\tilde{\varrho}}=\hat{D}^\dagger\hat{\varrho}\hat{D}$ in the displaced reference frame, which is formally given by
\begin{align}
	\frac{\partial}{\partial t}\hat{\tilde{\varrho}} &= \frac{1}{{\rm i}\hbar}[\hat H_{\rm eff}', \hat{\tilde{\varrho}}] + \sum_{n=1}^M \kappa_n\mathcal{D}[\hat{D}^\dagger\hat{a}_n\hat{D}]\hat{\tilde{\varrho}}\label{eq:displaced_master_equation}\\
    &\equiv   \mathcal{L}_\mathrm{D} \hat{\tilde{\varrho}}\,,\nonumber
\end{align}
where the explicit form of the superoperator $\mathcal{D}[\hat{X}]$, with $\hat X=\hat{D}^\dagger\hat{a}_n\hat{D}$, is given in Eq.\ \eqref{eq:Dissipator} and
\begin{equation}
    \label{eq:H_eff_formal}
    \hat{H}_{\mathrm{eff}}' = \hat{D}^\dagger\hat{H}_{\mathrm{eff}}\hat{D} - \ii \hbar \hat{D}^\dagger\frac{\partial \hat{D}}{\partial t}\,.
\end{equation}

Subsequently, we use the Nakajima-Mori-Zwanzig method~\cite{breuer2002theory}, for which we define the projector $\mathcal{P}$ acting on a generic operator $\hat{O}$ in the composite Hilbert space of cavity and motional degrees of freedom as the projection onto the target subspace in which the cavity fields are in the vacuum state: $\mathcal{P}\hat{O} = \bra{\mathrm{vac}}\hat{O}\ket{\mathrm{vac}} \otimes \ket{\mathrm{vac}}\bra{\mathrm{vac}}$. Correspondingly, the projector $\mathcal{Q} = 1-\mathcal{P}$ identifies the orthogonal subspace, and the master equation \eqref{eq:displaced_master_equation} can be decomposed into coupled equations for the projected density operators $\hat{v} = \mathcal{P}\hat{\tilde{\varrho}}$ and $\hat{w} = \mathcal{Q}\hat{\tilde{\varrho}}$~\cite{Gardiner:QNoise}:
\begin{align}
    \frac{\partial}{\partial t}\hat{v} &= \mathcal{P}\mathcal{L}_\mathrm{D}\hat{v} + \mathcal{P}\mathcal{L}_\mathrm{D}\hat{w}\label{Eq:eom_v}\,,\\
    \frac{\partial}{\partial t}\hat{w} &= \mathcal{Q}\mathcal{L}_\mathrm{D}\hat{v} + \mathcal{Q}\mathcal{L}_\mathrm{D}\hat{w}\label{Eq:eom_w}\,.
\end{align}
We assume that the initial state of the system, $\hat{\tilde{\varrho}}(t=0)$, lies within in the target subspace; that is, the cavity modes are in the vacuum state, and therefore $\mathcal{Q}\hat{\tilde{\varrho}}(t=0) = \hat{w}(t=0) = 0$. The condition that the cavity fields remain at all times in the vacuum state, $\hat{w}(t) = 0$, consists in requiring that~\cite{Jaeger:2022}
\begin{equation}
    \label{eq:condition}
    \mathcal{Q}\mathcal{L}_\mathrm{D}\hat{v}= 0\,,
\end{equation}
which ensures that no probability flow occurs from the vacuum subspace to states with non-zero photonic occupations. This condition allows us to determine the equation that $\hat{r}$, Eq.\ \eqref{eq:r_general}, must satisfy. In fact, Eq.\ \eqref{eq:condition} takes the explicit form
\begin{equation}
    \frac{1}{{\rm i}\hbar}\mathcal{Q}[\hat H_{\rm eff}', \hat{v}] + \sum_{n=1}^M \kappa_n\mathcal{Q}\mathcal{D}[\hat{D}^\dagger\hat{a}_n\hat{D}]\hat{v} = 0\,,\label{Eq:QLP=0}
\end{equation}
where, using the Baker-Campbell-Hausdorff formula~\cite{Gardiner:QNoise}, the transformed Hamiltonian $\hat H_{\rm eff}'$, Eq.\ \eqref{eq:H_eff_formal}, and the field operators $\hat{a}_n$ can be written as
\begin{align}
    \hat H_{\rm eff}' &= \sum_{l=0}^\infty \frac{(-1)^l}{l!}\left[\hat{r},\hat{H}_\mathrm{eff} - \frac{\ii \hbar}{l+1}\frac{\partial \hat{r}}{\partial t}\right]_l\,,\label{eq:Heff_commutators}\\
    \hat{D}^\dagger \hat{a}_n \hat{D} &= \sum_{l=0}^\infty \frac{(-1)^l}{l!}[\hat{r},\hat{a}_n]_l\,.
\end{align} 
Here, $[\hat{r},\hat{O}]_l$ denotes the $l$-order nested commutator, explicitly defined as $[\hat{r},\hat{O}]_0 = \hat{O}$, $[\hat{r},\hat{O}]_1 \equiv [\hat{r},[\hat{r},\hat{O}]_0]= [\hat{r},\hat{O}]$, and in general $[\hat{r},\hat{O}]_{l+1} = [\hat{r},[\hat{r},\hat{O}]_l]$. Equation \eqref{Eq:QLP=0} defines the time evolution of the operators $\hat{\alpha}_n$: 
\begin{equation}
    \label{eq:alpha:gen:0}
    \frac{\partial}{\partial t}\hat{\alpha}_n = \mathcal F_n(\{\hat{\alpha}_m\}, \hat{O}_{\rm at})\,,
\end{equation}
where $\mathcal{F}_n$ is a functional that depends on the operators $\hat{\alpha}_m$ ($m=1,\dots,M$) and other atomic operators $\hat{O}_{\rm at}$. Provided it exists, the solution $\hat{\alpha}_n(t)$ warrants that the dynamics remains confined to the target subspace at all times, so that Eq.\ \eqref{Eq:eom_v} reduces to
\begin{equation}
    \label{Eq:v}
   \frac{\partial}{\partial t}\hat{v}(t) = \mathcal{P}\mathcal{L}_{\mathrm{D}}\hat{v}(t)\,,
\end{equation}
where $\mathcal{L}_{\mathrm{D}}=\mathcal{L}_{\mathrm{D}}[\{\hat \alpha_n(t)\}]$. By tracing out the cavity degrees of freedom, Eq.\ \eqref{Eq:v} becomes an equation for the density operator of the atomic degrees of freedom:
\begin{equation}
    \label{Eq:mu}
    \frac{\partial}{\partial t}\hat{\mu}=\mathcal{L}_{\rm at}[\{\hat{\alpha}_n(t)\}]\hat{\mu}\,,
\end{equation}
where
\begin{equation}
    \hat{\mu}(t) = \bra{\text{vac}}\hat{v}(t)\ket{\text{vac}}\,.
\end{equation}
The dynamics is governed by the atomic superoperator $\mathcal{L}_{\rm at}$ that depends explicitly on the operators $\hat{\alpha}_n$. In general, $\hat{\alpha}_n$ has an integral form that depends on the history of the atomic state $\hat{\mu}$. As a consequence, Eq.\ \eqref{Eq:mu} is an integro-differential equation. Our goal is to identify regimes in which it is local in time and, in particular, takes the form of a Lindblad master equation. Achieving this requires additional assumptions, which we discuss next. We will show that these assumptions are motivated by existing experimental parameter regimes, and thus it makes the resulting Lindblad master equation a powerful framework for predicting experimental results and modeling the experimental dynamics.

\subsection{Lindblad Master Equation}
Before analyzing the regimes in which a Lindblad form can be attained, it is instructive to first discuss the meaning of Eq.\ \eqref{eq:alpha:gen:0}. These equations are a function of atomic operators in a mixed representation, in which both operators $\hat{\alpha}_n$ and the density operator $\hat{\mu}$ are evolved in time [see Eq.\ \eqref{Eq:mu}]. 

In order to unveil the time-dependence of the individual elements of the equation---and thereby acquire information about the properties of the solutions---it is useful to consider the equation for the expectation value $\langle\partial_t\hat{\alpha}_n\rangle = {\rm Tr}\{(\partial_t\hat{\alpha}_n(t)) \hat{\mu}(t)\}$. For the typical scenarios considered here, this can be cast in the form $\langle \partial_t\hat{\alpha}_n\rangle = h_n\langle\hat{\alpha}_n\rangle + \langle \hat{O}_{\rm at} \rangle$, where $h_n$ contains characteristic frequencies of the cavity modes and $\langle \hat{O}_{\rm at} \rangle$ evolves with the characteristic frequencies of the atomic system. Let $\tau_\mathrm{modes}$ denote the characteristic timescale of the cavity modes' dynamics, and $\tau_{\rm at}$ the characteristic timescales of the atomic motion. The latter depends on the motional energy and thus, loosely speaking, on the temperature according to the definition used in laser cooling~\cite{Wineland:1979, Stenholm:1986, Domokos:2003}. When $\tau_{\rm modes}\ll\tau_{\rm at}$, a coarse-graining timescale $\Delta t$ can be identified such that $\tau_{\rm modes} \ll  \Delta t \ll \tau_{\rm at}$. A coarse-graining procedure can then be applied to the equation of motion, which amounts to performing a time averaging of the coupled equations for the atomic density operator $\hat{\mu}(t)$ and the atomic operator $\hat{\alpha}_n(t)$:
\begin{equation}
    \overline{\hat{\alpha}_n(t)} = \frac{1}{\Delta t}\int_t^{t + \Delta t} \mathrm{d}\tau~\hat{\alpha}_n(\tau)\,.\label{eq:time_average_alpha}
\end{equation}
According to this procedure, the time-averaged derivative $\langle\overline{\partial_t \hat{\alpha}_n(t)}\rangle=0$ vanishes in leading order, and the time derivatives in Eq.\ \eqref{eq:Heff_commutators} are effectively averaged out. As a result, the operators $\hat{\alpha}_n$ entering Eq.\ \eqref{Eq:mu} become local in time, and the master equation acquires Lindblad form. 

In what follows, we discuss two relevant examples in which the atom-only Lindblad master equation can be derived. These correspond to two limiting regimes: the first is the \textit{weak-coupling limit}, characterized by vanishing intracavity photon numbers, and the second relies on an \textit{adiabatic expansion}, applicable when the cavity degrees of freedom evolve faster than the atomic motional ones. This master equation is one of the main results of this work. In fact, differing from other perturbative expansions~\cite{Breuer:2001}, the form of our adiabatic master equation permits us to include non-adiabatic corrections in a systematic manner without violating positivity. For this reason, it remains valid across a broad range of optomechanical coupling strengths, from perturbative cavity-mediated interactions to those that dominate the atomic motion.

\section{Lindblad master equation in the weak-coupling limit}
\label{Sec:weak}

The formalism described above leads to a relatively simple and compact master equation of Lindblad form in the limit in which the coupling between atomic motion and cavity modes is sufficiently weak. This equation is valid when the intracavity photon number remains close to zero in the laboratory frame, and its regime of validity is qualitatively illustrated in Fig.~\ref{fig:1}(d). We highlight that, although a master equation under similar assumptions has been derived in Ref.~\cite{Nimmrichter:2010}, the master equation developed in this section is of Lindblad form and therefore preserves the positivity of the density operator.

In what follows, we outline the steps required to explicitly derive the atom-only Lindblad master equation and quantify its regime of validity. We then benchmark the predictions of the atom-only master equation against those of the optomechanical master equation from which it has been derived. To this end, we choose specific parameters for which the cavity-assisted dynamics leads to cooling of the atomic motion. Finally, we illustrate the power of the formalism by deriving an analytical description of the cooling dynamics. 

\subsection{Derivation of the Atom-Only Master Equation in the Weak-Coupling Limit}

Before we start, we introduce the eigenbasis $\{\ket{\beta}\}$ of the mechanical Hamiltonian $\hat{H}_{\rm S}$, satisfying the eigenvalue equation $\hat H_{\rm S}\ket{\beta}=\mathcal E_\beta\ket{\beta}$ with eigenenergies $\mathcal{E}_\beta$. By virtue of the completeness relation, we express the operator $\hat\alpha_n$ in this basis as
\begin{equation}
    \label{eq:decomp_alpha_eps_beta}
    \hat\alpha_n=\sum_{\beta,\beta'}\alpha_{\beta,\beta'}^{(n)}\ket{\beta}\bra{\beta'}\,.
\end{equation}
The coefficients $\alpha_{\beta,\beta'}^{(n)}=\bra{\beta}\hat\alpha_n\ket{\beta'}$ obey a set of differential equations that are derived from Eq.\ \eqref{Eq:QLP=0}. These equations can be simplified when $\langle \hat{\alpha}_n^\dagger \hat{\alpha}_n \rangle \ll 1$. This condition corresponds to a vanishing photon number in the laboratory frame; in fact, $\langle \hat{\alpha}_n^\dagger \hat{\alpha}_n \rangle$ in the transformed reference frame corresponds to the mean intracavity photon number $\langle \hat{a}_n^\dagger \hat{a}_n \rangle$ in the laboratory frame. Under this assumption, we truncate the nested commutators in Eq.\ \eqref{Eq:QLP=0} to first order and obtain a set of first-order differential equations for the coefficients $\alpha_{\beta,\beta'}^{(n)}$. We give their form for a single cavity mode ($M=1$):
\begin{equation}
    \begin{aligned}
        \label{eq:alpha:gen}
    {\rm i}\frac{\partial}{\partial t}\alpha_{\beta,\beta'} &= \left(\frac{\mathcal{E}_\beta - \mathcal{E}_{\beta'}}{\hbar} - \tilde{\Delta} - {\rm i} \kappa\right)\alpha_{\beta,\beta'}\\
    & + U\sum_{\beta''\neq\beta}\bra{\beta}\hat{\Theta}_{11}\ket{\beta''}\alpha_{\beta'',\beta'}\\
    & + \eta\bra{\beta}\hat{\Theta}_{1\mathrm{p}}\ket{\beta'}+  \delta_{\beta,\beta'}\zeta e^{\ii \delta \omega_\mathrm{p} t}\,,
    \end{aligned}
\end{equation}
where we neglected all terms of second order in $\alpha_{\beta,\beta'}$ or higher. In writing Eq.\ \eqref{eq:alpha:gen} we dropped the superscript $n$. Here, $\tilde{\Delta} = \Delta - U\bra{\beta}\hat{\Theta}_{11}\ket{\beta}$ denotes the effective cavity detuning, which includes the atomic-induced dispersive shift $U\bra{\beta}\hat{\Theta}_{11}\ket{\beta}$.

The assumption of vanishing intracavity photons imposes constraints on the pump strengths: for a longitudinal pump, it requires $|\zeta|\ll |\tilde{\Delta}+\ii\kappa|$; for a transverse pump, it requires $|\eta|N\ll |\tilde{\Delta}-\delta+\ii\kappa|$, where $\delta$ represents the energy difference between the coupled states~\cite{Schuetz:2015}; its explicit form depends on the Hamiltonian $\hat{H}_\mathrm{S}$. 

\subsection{Dynamics of an Atom in a Transversely-Pumped Resonator}

We now provide the explicit form of $\hat{\alpha}$ in the single-mode case (with the index $n$ dropped) for insertion into the master equation. To simplify the notation, we consider the case of a single atom whose motion is restricted to the cavity axis of a standing-wave resonator with wave number $k$. Let $\hat{H}_\mathrm{S}=\hat{p}^2/(2m)$ and $\hat{\Theta}_{11}=\cos^2(k\hat{x})$. The resonator is driven by a transverse pump that is orthogonal to the cavity axis. Accordingly, in Eq.\ \eqref{eq:alpha:gen}, we set $\zeta=0$ and identify $\hat\Theta_{1\mathrm{p}}=\cos(k\hat x)$. 

In these settings, the eigenstates of the Hamiltonian $\hat{H}_\mathrm{S}$ are the momentum eigenstates $\ket{p}$, with corresponding eigenenergies $\mathcal{E}_p = p^2/(2m)$. The matrix elements of $\hat{\Theta}_{1\mathrm{p}} = \cos(k\hat{x})$ are $\bra{p'}\hat{\Theta}_{1\mathrm{p}}\ket{p} = (\delta_{p',p-\hbar k} + \delta_{p',p+\hbar k})/2$, and thus couple states with an energy difference $\mathcal{E}_{p \pm \hbar k} - \mathcal{E}_{p} = \pm \hbar k p/m + \hbar \omega_\mathrm{R}$, where $\omega_{\rm R}=\hbar k^2/(2m)$ is the recoil frequency. Applying the coarse-graining procedure described above, we set the time derivative in Eq.\ \eqref{eq:alpha:gen} to zero and obtain with Eq.\ \eqref{eq:decomp_alpha_eps_beta} an explicit expression for the operator $\hat{\alpha}$: 
\begin{equation}
    \label{alpha_op:p}
    \hat{\alpha} = \sum_p(\alpha_-(p)\ket{p - \hbar k}\bra{p} + \alpha_+(p)\ket{p + \hbar k}\bra{p})\,,
\end{equation}
where the coefficients are
\begin{equation}
\label{alpha:p}
    \alpha_\pm(p) = \frac{\eta/2}{\Delta \mp k p/m - \omega_\mathrm{R} + \ii \kappa}\,.
\end{equation}
Note that we neglected here the shift $U$ assuming that $\kappa\gg |U|$, see Appendix \ref{App:Cavity cooling} for details. 

The effective master equation is obtained by inserting $\hat{\alpha}$ from Eq.\ \eqref{alpha_op:p} in Eq.\ \eqref{Eq:mu}. We report here the equations for the diagonal elements $\Pi_p = \bra{p}\hat{\mu}\ket{p}$, which represent the probability that the atom occupies the momentum state $\ket{p}$:
\begin{align}
    \frac{\partial}{\partial t}\Pi_p =& -(r_-(p) + r_+(p))\,\Pi_p + r_+(p-\hbar k)\,\Pi_{p-\hbar k}\nonumber\\
                                        &+ r_-(p+\hbar k)\,\Pi_{p+\hbar k}\,,\label{eq:P}
\end{align}
where $r_\pm(p) = 2\kappa |\alpha_\pm(p)|^2$ are the scattering rates for processes that increase (+) or decrease (-) the momentum. We have neglected the coupling to the coherences $\bra{p'}\hat{\mu}\ket{p}$ with $p' \neq p$, as these terms correspond to processes of higher order in the small parameter $|\eta|/\kappa$. Further details are provided in Appendix~\ref{App:Cavity cooling}. The set of equations \eqref{eq:P} admits a steady-state solution for $\Delta<0$. As shown in Appendix \ref{App:Cavity cooling:Steady-state distribution}, the steady-state distribution takes the form of a $q$-Gaussian with power-law tails, see also Ref.\ \cite{Niedenzu:2011} for a semiclassical treatment. The steady-state distribution approaches a Gaussian for $|\Delta|\gg\omega_\mathrm{R}$~\cite{Niedenzu:2011,Schuetz:2013}.

\subsection{Benchmark: Cavity Cooling}

In the weak-coupling limit, the timescale separation is warranted, since the characteristic timescale at which the atomic motion is modified scales with $\eta\sqrt{N}$. Moreover, the average photon number in the laboratory frame remains small at all times. This regime includes, for example, cavity-assisted cooling via coherent photon scattering, see Refs.\ \cite{Vuletic:2000,Morigi:2007}. In the following, we focus on this regime and analyze the validity of the atom-only master equation in describing the cooling dynamics of the atomic motion. The explicit expressions of the optomechanical and the atom-only master equations are reported in Appendix~\ref{App:Cavity cooling}.

Figure~\ref{fig:cavity_cooling} compares the time evolution of the mean kinetic energy as predicted by the optomechanical and atom-only master equations. For the chosen parameter sets---within the regime where the atom-only master equation is expected to be valid---the predictions of both models show excellent agreement. 

We now further elaborate on the model of Eq.~\eqref{eq:P}. A relatively simple way to obtain an analytical prediction for the dynamics is to make the ansatz $\Pi_p(t) = \exp(-\beta(t)p^2/(2m))/Z$, with $Z = \int_{\mathbb{R}}{\rm d}p~\exp(-\beta(t) p^2/(2m))$. This corresponds to assuming that the momentum distribution is a Gaussian at all times, with a time-dependent inverse temperature $\beta(t)= 1/(2E_\mathrm{kin}(t))$~\cite{Wineland:1979,Stenholm:1986,Morigi:2001}. Through this ansatz, $\beta(t)$ is directly related to the mean kinetic energy, $E_\mathrm{kin}(t) = \int_{\mathbb{R}}\mathrm{d}p~p^2/(2m)~\Pi_p(t)$. The equation for $E_\mathrm{kin}$ is obtained by multiplying  Eq.\ \eqref{eq:P} by $p^2/(2m)$ and integrating over $p$:
\begin{equation}
    \frac{\partial}{\partial t}E_\mathrm{kin} = -\gamma E_\mathrm{kin} + h\label{eq:eom_E_kin}\,,
\end{equation}
where the coefficients are (here reported to leading order in $\omega_\mathrm{R}$)
\begin{equation}
    \begin{aligned}
            \gamma &= \omega_\mathrm{R}|\eta|^2\frac{-8\Delta\kappa}{\left[\Delta^2 + \kappa^2\right]^2}\,,\\
            h &= \omega_\mathrm{R}|\eta|^2\frac{\hbar \kappa }{\Delta^2 + \kappa^2}\,.
    \end{aligned}
\end{equation}
For $\gamma > 0$ (thus $\Delta < 0$), the equation has a steady-state solution, which is approached by an exponential decay at rate $\gamma$. The stationary distribution is a Gaussian with mean kinetic energy
\begin{equation}
    E_\mathrm{kin}^\mathrm{(ss)} = \frac{h}{\gamma}=\frac{\hbar[\Delta^2 + \kappa^2]}{8|\Delta|}\,,
\end{equation}
which reaches the minimum $E_\mathrm{kin}^\mathrm{(ss)}=\hbar\kappa/4$ at $\Delta = -\kappa$~\cite{Domokos:2003,Schuetz:2013}. This result has also been obtained using the Keldysh formalism~\cite{Piazza:2014}. In Fig.\ \ref{fig:cavity_cooling}, the dotted lines display the predictions of Eq.\ \eqref{eq:eom_E_kin} based on the Gaussian ansatz: the curves reproduce the exponential decay of the mean kinetic energy, even though they may not quantitatively reproduce the cooling rate or the steady-state value.

We test the validity of the Gaussian ansatz by determining the Kurtosis 
\begin{equation}
    \label{Eq:Kurtosis}
    \mathcal{K}=\langle(\delta\hat{p})^4\rangle/\langle(\delta\hat{p})^2\rangle^2\,,
\end{equation} 
with $\delta\hat{p} = \hat{p} - \langle\hat{p}\rangle$. The insets in Fig.\ \ref{fig:cavity_cooling} display the Kurtosis of the distribution evaluated using the optomechanical master equation. The deviation of the Kurtosis from the value 3 signals a deviation from Gaussianity~\cite{Chissom:1970}. For both considered parameter sets, we observe, indeed, such deviations and, correspondingly, also discrepancies between the simplified model that uses a Gaussian approximation and the numerical results. Remarkably, both approaches, the optomechanical and the atom-only master equations, capture this non-Gaussian behavior.

\begin{figure}[!ht]
    \centering
    \includegraphics[scale = 1]{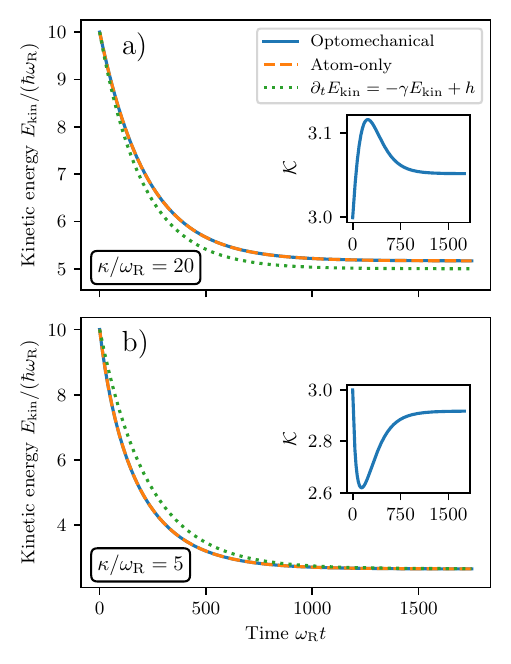}
    \caption{Evolution of the mean kinetic energy $E_\mathrm{kin}$ for an atom undergoing cavity cooling. Energy and time are in units of $\hbar \omega_\mathrm{R}$ and of $\omega_\mathrm{R}^{-1}$, respectively. The solid blue and the dashed orange curves are, respectively, obtained by numerically integrating the optomechanical, Eq.\ \eqref{eq:opto_master_equation_disp}, and the atom-only master equation, Eq.\ \eqref{Eq:mu}; see Appendix\ \ref{App:Cavity cooling} for their explicit expressions. The dotted green curve corresponds to the prediction of the Gaussian ansatz, Eq.\ \eqref{eq:eom_E_kin}. The parameters are: a) $(\eta,\Delta,\kappa) = (1, -20, 20)\,\omega_\mathrm{R}$ and b) $(\eta,\Delta,\kappa) = (1, -20, 5)\,\omega_\mathrm{R}$.  The initial atomic state is a thermal distribution $\exp(-\beta \hat{p}^2/(2m))$ with initial temperature $\beta^{-1} = 20\,\hbar\omega_\mathrm{R}$. In the optomechanical master equation, the cavity is initially in the vacuum state $\ket{\mathrm{vac}}$. The insets show the Kurtosis, Eq.\ \eqref{Eq:Kurtosis}, extracted from the full simulations of the optomechanical master equation. All equations were implemented and numerically integrated using the library described in Ref.~\cite{Kramer:2018}.}
    \label{fig:cavity_cooling}
\end{figure}

\subsection{Discussion}

First, we mention that the generalization of the atom-only master equation to $N$ atoms is straightforward, as we simply need to exchange both operators $\hat{\Theta}_{1\mathrm{p}} = \sum_{j=1}^N\cos(k\hat{x}_j)$ and $\hat{\Theta}_{11}=\sum_{j=1}^N\cos^2(k\hat{x}_j)$.  In this case, typical cooling rates are rescaled by $N$ and superradiant effects can accelerate the cooling into the stationary state~\cite{Salzburger:2009}. 

We note that the treatment in this section requires that the number of intracavity photons is small at all times. This condition can be relaxed for trapped atoms in the Lamb-Dicke regime, when the size of the atomic wave packet is smaller than the laser wavelength, see, e.g.,\ Ref.~\cite{Zippilli:2005}. The addition of interparticle interactions can also be treated within this formalism by choosing an adequate basis of many-body states and identifying accordingly the timescales that allow for a coarse-graining, see, e.g., Refs.~\cite{Fogarty:2016,Cormick:2012}. The low-photon number approximation, in this section, is equivalent to the weak-coupling assumption. The latter is the approximation at the basis of the atom-only master equation of Ref.~\cite{Jaeger:2022}. In general, the applicability of this approach is questionable for describing the formation of correlated atomic states, such as superradiant atomic configurations, where the magnitude of the cavity-mediated potentials is comparable to the characteristic energies of the atomic motion. In the next section, we resolve this issue with a systematic approach that is also valid for strong cavity-mediated interactions.

\section{Lindblad master equation in the adiabatic limit and beyond}
\label{Sec:strong}

Our goal is to describe the regime of strong cavity-mediated interactions, in which collective phenomena emerge, such as self-organization~\cite{Mivehvar2021Cavity}, the Dicke phase transition of CQED~\cite{Baumann2010Dicke}, and time crystals~\cite{Kessler:2021,Kongkhambut:2022}, to name some prominent examples. This regime is typically accompanied by relatively large intracavity photon numbers, rendering perturbative treatments of the coupling with the cavity modes inadequate. While mean-field descriptions go beyond the perturbative regime and have been successfully employed to study such phenomena, they inherently neglect cavity-induced atom-atom correlations, that are essential for describing correlated phases of matter~\cite{Lin:2019,Klinder:2015, Halati:2025a, Halati:2025b}. The atom-only master equation developed in this work overcomes these limitations, thus enabling the study of genuine quantum dynamics and offering the possibility to derive simplified models that are amenable to analytic treatments and more efficient numerical treatments. It extends the parameter regime of validity and covers the area  of Fig.~\ref{fig:1}(d). 

In this section, we first derive the atom-only master equation based on a time-scale separation, that permits us to implement an adiabatic expansion; the details of the derivation are provided in Appendix~\ref{App:A}. Then, we test its predictions for the Bose-Hubbard model of cavity QED by a systematic comparison with the full optomechanical model.  

\subsection{Adiabatic Limit}

To derive the master equation, we first consider once again the extreme limit of particles with infinite mass, where the kinetic energy is zero, and an exact form can be derived even for large intracavity photon numbers. In fact, after discarding the kinetic energy, all atomic operators depend solely on the atomic positions and can be treated as scalars. Thus, the Hamiltonian $\hat{H}_{\mathrm{eff}}$, Eq.\ \eqref{eq:def_H_eff}, is diagonal in the particles’ position basis and the photonic sector can be brought to a diagonal form using a displacement transformation, Eq.\ \eqref{eq:D}, generated by
\begin{align}
    \hat{r}(t)=\sum_{n=1}^{M}(\hat{a}_n^\dagger\hat{\alpha}_{0,n}(t)-\hat{\alpha}_{0,n}^\dagger(t)\hat{a}_n)\,,\label{eq:r}
\end{align} 
where the atomic operators $\hat{\alpha}_{0,n}$ satisfy the equations of motion 
\begin{eqnarray}
	\frac{\partial}{\partial t}\hat{\alpha}_{0,n} &=& \sum_{m=1}^M\hat{\chi}_{nm}\hat{\alpha}_{0,m}-\ii\hat{G}_{n}(t) \label{eq:easy}\,,
\end{eqnarray}
with
\begin{align}
    \hat{\chi}_{nm}=-\ii\left(\hat{\Omega}_{nm}-\ii\kappa_n\delta_{n,m}\right)\,.\label{eq:chi}
\end{align}
For small contributions of the kinetic energy (large mass and/or small temperatures), this is still a good solution for the dynamics of particles in a resonator, as long as the diabatic effects due to the atomic motion are negligible. In this case, the master equation for the atomic density operator $\hat{\mu}$ takes the form
\begin{equation}
	\frac{\partial}{\partial t}\hat{\mu} = \frac{1}{{\rm i}\hbar}[\hat{H}_{\mathrm{eff},0}^{\rm at},\hat\mu] + \sum_{n=1}^M \kappa_n \mathcal{D}[\hat{\alpha}_{0,n}]\hat\mu\label{eq:eff_master_equation:0}\,,
\end{equation}
with
\begin{equation}
	\hat{H}_{\mathrm{eff},0}^{\rm at} = \hat{H}_\mathrm{S}+ \frac{\hbar}{2}\sum_{n=1}^M \left(\hat{G}_n^\dagger\hat{\alpha}_{0,n} + \hat{\alpha}_{0,n}^\dagger \hat{G}_n\right)\label{eq:H_eff:0}\,.
\end{equation}
This master equation is of Lindblad form and includes a cavity-mediated time-dependent potential [see second term on the RHS of Eq.~\eqref{eq:H_eff:0}] and dissipation with time-dependent jump operators [see the second term on the RHS of Eq.~\eqref{eq:eff_master_equation:0}].

To estimate the order of magnitude of the diabatic corrections, we consider the first nontrivial term ($l=1$) of the nested commutators in Eq.\ \eqref{eq:Heff_commutators} that originates from the kinetic energy $\hat{H}_{\rm kin} = \sum_{j}\hat{\vec{p}}_j^2/(2m)$:
\begin{align}
    \hat{H}_{\mathrm{pert}}=\sum_{n=1}^{M}([\hat{\alpha}_{0,n}^{\dagger},\hat{H}_{\rm kin}]\hat{a}_n+\hat{a}_n^{\dagger}[\hat{H}_{\rm kin},\hat{\alpha}_{0,n}])\,,\label{eq:pert}
\end{align}
where we used that $\hat{\alpha}_{0,n}$ depends solely on the atoms' positions, so that $[\hat{H}_\mathrm{S},\hat{\alpha}_{0,n}] = [\hat{H}_{\rm kin},\hat{\alpha}_{0,n}]$. This term constitutes the first non-vanishing diabatic correction coupling the photonic vacuum to states with nonzero intracavity photon number. We quantify its effective strength by evaluating the expectation value of the commutator $[\hat{H}_{\rm kin},\hat{\alpha}_{0,n}]$ over the atomic state. In order to provide an estimate, we take a single-mode cavity ($M=1$) without longitudinal pump ($\zeta=0$) and approximate $\hat{\Omega}_{nn} \approx -\Delta_n$. We further assume that the atomic state is characterized by a Gaussian momentum distribution, where the single-particle width is denoted as $\Delta p$. Therefore, the mean kinetic energy per particle is $E_{\rm kin}=\Delta p^2/(2m)$. With this, one can estimate
$$
    \Vert[\hat{H}_{\rm kin},\hat{\alpha}_{0,n}]\Vert \sim \sqrt{4E_{\rm kin}\hbar\omega_\mathrm{R}} \frac{|\eta_n\sqrt{N}|}{|\Delta_n + \ii\kappa_n|}\,.
$$
Extending the adiabatic theorem to this situation, this strength shall be compared with  the cavity frequency and decay rate $| \Delta_{n} + \ii\kappa_n|$, giving the condition for adiabatic dynamics: 
\begin{align}
    \frac{|\eta_n\sqrt{N}|}{\sqrt{\Delta_n^2+\kappa_n^2}}\sqrt{\frac{4E_{\rm kin}\omega_\mathrm{R}}{\hbar(\Delta_n^2+\kappa_n^2)}} \ll 1\,,\label{eq:cond_pert_kin_energy}
\end{align}
with $E_{\rm kin}\ge \hbar\omega_\mathrm{R}$. The adiabatic condition~\eqref{eq:cond_pert_kin_energy} therefore implies that the effective master equation \eqref{eq:eff_master_equation:0} can be valid even in rather strong-coupling regimes, where $\eta_n\sqrt{N}\gtrsim |\Delta_n+\ii\kappa_n|$, provided the atomic ensemble is sufficiently cold that $|\Delta_n+\ii\kappa_n|^2\gg E_{\rm kin}\omega_\mathrm{R}/\hbar$. 

Condition~\eqref{eq:cond_pert_kin_energy} also implies that the master equation~\eqref{eq:eff_master_equation:0} is valid only for a finite time. In fact, the Lindblad terms $\mathcal{D}[\hat{\alpha}_{0,n}]$ solely depend on the atomic positions. They emerge from incoherent scattering processes, where photons are emitted by the resonators. The dynamics they describe is a projective measurement of the atoms' positions, thus effectively causing heating of the quantum gas over time~\cite{Sierant:2019, Halati:2020a}. We can estimate the corresponding heating rate by examining the factor scaling the Lindblad terms. For a single cavity mode, the rate is
\begin{equation}
    \Gamma_n \simeq \frac{\kappa_n |\eta_n|^2}{\Delta_n^2 + \kappa_n^2}\,,
\end{equation}
where we have neglected the dispersive shift. Hence, the master equation~\eqref{eq:eff_master_equation:0} is expected to deliver a reliable description for times $t\lesssim 1/\Gamma_n$. We can extend the timescale over which the master equation is valid by including the first diabatic corrections. This is discussed in the next section.

\subsection{Diabatic corrections}
\label{sec:alpha1}
We now extend the master equation by taking into account diabatic effects in leading order. For this purpose, we write 
\begin{align}
	\hat{\alpha}_n= \hat{\alpha}_{0,n} + \hat{\alpha}_{1,n}\,,
\end{align}
where $\hat{\alpha}_{0,n}$ is the solution of Eq.~\eqref{eq:easy} that depends solely on the atoms' positions, and $\hat{\alpha}_{1,n}$ incorporates the atoms' momenta by taking into account the term described in Eq.~\eqref{eq:pert}. Before we delve into the derivation, based on a perturbative analysis, it is important to identify a reasonable perturbation order. Equation~\eqref{eq:H_eff:0} contains two terms, namely $\hat{H}_\mathrm{S}$ and the cavity-mediated term $\propto\hat{\alpha}_{0,n}$. We want their effects on the dynamics of the atoms to be comparable, yet we also want these effects to occur on a much slower timescale than the typical timescale of the cavity degrees of freedom, which is set by $|\Delta_n+\ii\kappa_n|$. Accordingly, we identify the two small parameters
\begin{align}
    \varepsilon_1\equiv \frac{\sqrt{4E_{\rm kin}\omega_\mathrm{R}/\hbar}}{|\Delta_n + \ii\kappa_n|},\quad\varepsilon_2\equiv\frac{|\eta_n|\sqrt{N}}{|\Delta_n + \ii\kappa_n|}\label{Eq:pert_parameter_eps}\,,
\end{align}
where $\varepsilon_1$ and $\varepsilon_2$ represent the contributions of the kinetic energy and the term $\propto\hat{\alpha}_{0,n}$ in Eq.~\eqref{eq:H_eff:0}, respectively. Note that condition \eqref{eq:cond_pert_kin_energy} corresponds to assuming $\varepsilon_1\varepsilon_2\ll 1$. However, the two terms appear also individually, in fact $\hat{H}_\mathrm{S}/(\hbar|\Delta_n + \ii \kappa_n|) \sim \varepsilon_1$ and $\hat{\alpha}_{0,n} \sim \varepsilon_2$. In the rest of this section, we will assume parameters for which $\varepsilon_1\sim\varepsilon_2^2 \equiv \varepsilon^2$. This is chosen to describe the crossover regime from weak to strong interactions, where the energies of the cavity-mediated interactions ($\sim \varepsilon_2^2$) become comparable to the kinetic energy scaling ($\sim\varepsilon_1$). Our purpose is now to determine the correction $\hat{\alpha}_{1,n}\sim\varepsilon^3$, thereby pushing the validity of the master equation \eqref{eq:eff_master_equation:0} to include the lowest-order diabatic corrections. 

In Appendix \ref{App:A}, we show that the operators $\hat{\alpha}_{1,n}$ shall obey the equations of motion 
\begin{equation}
	\frac{\partial}{\partial t}\hat{\alpha}_{1,n} = \frac{1}{\ii\hbar}[\hat{H}_\mathrm{S},\hat{\alpha}_{0,n}] +\sum_{m=1}^M  \hat{\chi}_{nm}\hat{\alpha}_{1,m} \label{eq:harder}\,.
\end{equation}
After applying the coarse-graining averaging, we find that the fields $\hat{\vec{\alpha}}_\nu = (\hat{\alpha}_{\nu,1},\dots,\hat{\alpha}_{\nu,M})^T$ ($\nu = 0,1$) take the form
\begin{align}
    \hat{\vec{\alpha}}_0=&\underline{\underline{\hat{\chi}}}^{-1}\hat{\vec{b}}\,,\label{Eq:ss_alpha_0}\\
    \hat{\vec{\alpha}}_1=&-\underline{\underline{\hat{\chi}}}^{-1}\left(\frac{1}{\mathrm{i}\hbar}\left[\hat{H}_{\rm S},\hat{\vec{\alpha}}_0\right]\right)\,,\label{Eq:ss_alpha_1}
\end{align}
where $\underline{\underline{\hat{\chi}}}$ is the matrix with elements given in Eq.\ \eqref{eq:chi} and $\hat{\vec{b}}(t) = (\ii\hat{G}_1(t),\dots,\ii\hat{G}_M(t))^T$.
The master equation, including the lowest-order diabatic corrections, reads
\begin{eqnarray}
	\frac{\partial}{\partial t}\hat{\mu} 
	=\frac{1}{{\rm i}\hbar}[\hat{H}_\mathrm{eff}^{\rm at},\hat\mu] + \sum_{n=1}^M \kappa_n \mathcal{D}[\hat{\alpha}_n]\hat\mu\label{eq:eff_master_equation:1}\,,
\end{eqnarray}
with
\begin{equation}
	\hat{H}_\mathrm{eff}^{\rm at} = \hat{H}_{\rm S} + \frac{\hbar}{2}\sum_{n=1}^M \left(\hat{\alpha}_n^\dagger \hat{G}_n + \hat{G}_n^\dagger\hat{\alpha}_n\right)\label{eq:H_efffinal}\,.
\end{equation}
The Lindblad master equation \eqref{eq:eff_master_equation:1} permits to describe the onset of macroscopic correlations induced by the cavity fields and thus to characterize critical behavior in many-body CQED. In the next section, we verify this claim for a paradigmatic model: the Bose-Hubbard model of CQED.

\subsection{Extended Bose-Hubbard model}
\label{subsec:strong_coupling_Bose-Hubbard}

In the following, we test the predictions of the atom-only master equation for tightly bound bosons coupled to the field of an optical resonator. We derive the extended Bose-Hubbard model and make a systematic comparison with the full optomechanical model. This comparison is performed by determining the spectra of the optomechanical and of the atom-only Lindblad operators. It allows us to verify the regime of validity of our description, where the atom-only master equation provides an exact solution of the dynamics. 

To perform this validation, we consider a single cavity mode with photon annihilation operator $\hat{a}$, and assume that the atoms are bosons confined in an optical lattice, within the regime of the tight-binding and single-band approximation. As shown in Appendix \ref{App:C}, the Hamiltonian $\hat{H}_\mathrm{S}$ takes the form
\begin{equation}    
    \label{Eq:H:BH}
    \hat{H}_\mathrm{S} = - \hbar J \sum_{j=1}^{L-1}\left(\hat{b}_j^\dagger \hat{b}_{j+1} + \hat{b}_{j+1}^\dagger \hat{b}_{j}\right) + \hbar\frac{u}{2}\sum_{j=1}^{L}\hat{n}_j(\hat{n}_j - 1)\,,
\end{equation}
where $\hat b_j,\hat b_j^\dagger$ annihilate and create, respectively, a boson at site $j=1,\ldots,L$, with $[\hat b_i,\hat b_j^\dagger]=\delta_{i,j}$, $[\hat b_i,\hat b_j]=0$, and $\hat n_j=\hat b_j^\dagger\hat b_j$ counts the number of particles at site $j$. The coefficient $J$ is the hopping amplitude, and $u>0$ is the onsite repulsion.

As for the cavity field, we take large bare cavity detunings, such that the contribution from the dispersive shift can be neglected, $\Delta - U_{11}\hat{\Theta}_{11} \approx \Delta$. Moreover, as demonstrated in Appendix \ref{App:C}, for the tightly-bounded particles, the operator $\hat \Theta_{1\mathrm{p}}$ can be approximated by $\hat\Theta_{\rm BH}$, defined as
\begin{equation}
    \hat{\Theta}_{\rm BH}=\sum_{j=1}^{L-1} \left(Z_j \hat n_j + Y_j\left(\hat{b}_j^\dagger \hat{b}_{j+1} + \hat{b}_{j+1}^\dagger \hat{b}_{j}\right)\right)\,,
\label{Eq:Theta:BH_0}
\end{equation}
where the coefficients $Z_j,Y_j$ depend on the spatial mode functions of pump and cavity~\cite{Habibian:2013,Niederle:2016,Chanda:2022}. In what follows, we assume for simplicity $Y_j\approx0$ and $Z_j \approx (-1)^j$, which is reasonable within the tight-binding approximation when the wavelength of the optical lattice matches that of the cavity field and is the configuration of Refs.\ \cite{Landig:2016,Chanda:2022}. Under these conditions, $\hat\Theta_{\rm BH}\approx \sum_j(-1)^j\hat n_j$ [see Appendix~\ref{App:C}].

\begin{figure*}[!ht]
    \centering
    \includegraphics[width = 0.9\linewidth]{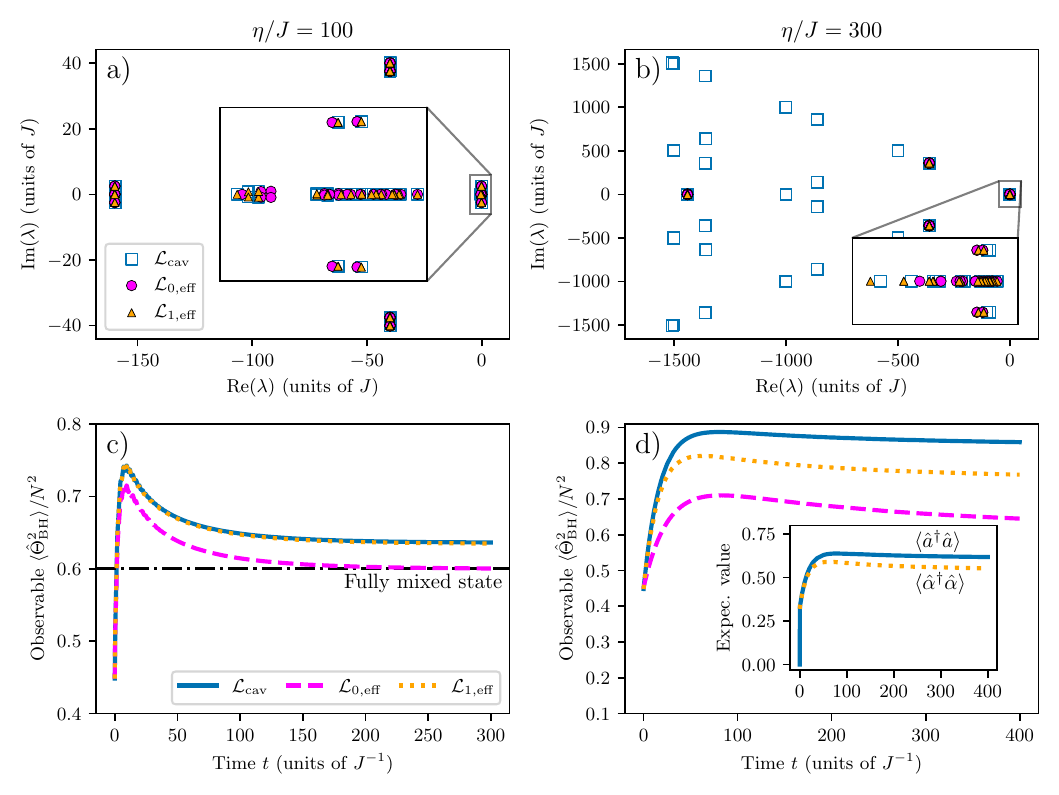}
    \caption{Benchmark of the atom-only master equation for the Bose-Hubbard model coupled to a dissipative cavity mode for a lattice with $N = 2$ bosons and $L = 4$ sites and open boundary conditions. The left panels correspond to the adiabatic regime, with $\varepsilon =0.04$. In the right panels, $\varepsilon =0.36$, such that diabatic corrections become relevant.
    Upper row: Eigenvalues $\lambda$ (in units of $J$) of the adiabatic atom-only Lindbladian $\mathcal{L}_{0,\mathrm{eff}}$, and of the atom-only Lindbladian $\mathcal{L}_{1,\mathrm{eff}}$ including the first diabatic correction. The eigenvalues of the optomechanical Lindbladian $\mathcal{L}_\mathrm{cav}$, Eq.\ \eqref{Eq:L:full}, are presented up to a chosen energy cutoff. The insets zoom into the slowest-decaying bundle of eigenvalues.
   Lower row: Evolution of $\langle \hat{\Theta}_{\rm BH}^2 \rangle$ predicted using the optomechanical (blue solid line) and the atom-only Lindbladians, with (orange dotted line) and without (pink dashed line) the first diabatic correction. The black dotted-dashed line in subplot c) marks the value $\langle \hat{\Theta}_\mathrm{BH}^2\rangle/N^2 = 0.6$ for a fully mixed state. The inset of subplot d) shows the dynamics of the cavity population $\langle \hat{a}^\dagger\hat{a}\rangle$ using the optomechanical Lindbladian (blue solid line) and the expectation value $\langle \hat{\alpha}^\dagger\hat{\alpha}\rangle$ for $\mathcal{L}_{1,\mathrm{eff}}$ (orange dotted line). The atoms are initially in the ground state of the Bose-Hubbard model $\hat{H}_\mathrm{S}$, Eq.\ \eqref{Eq:H:BH}. In the optomechanical simulations, the cavity mode is initially in the vacuum state. The parameters are $u = 2.5J$, $\Delta = -500J$, $\kappa = 500J$ and (a,c) $\eta=100J$, (b,d) $\eta=300J$. Numerical simulations were performed using the framework of Ref.~\cite{Kramer:2018}.}
    \label{fig:comparison_BH_ED}
\end{figure*}

Within this model, the optomechanical master equation \eqref{eq:opto_master_equation_disp} takes the form $\partial_t \hat{\varrho}=\mathcal L_{\rm cav}\hat{\varrho}$ with
\begin{equation}
    \mathcal L_{\rm cav}\hat{\varrho}=\frac{1}{{\rm i}\hbar}[\hat{H}_\mathrm{S} + \hat{H}_\mathrm{cav},\hat{\varrho}]+\kappa\mathcal{D}[\hat a]\hat{\varrho}\,,
\label{Eq:L:full}
\end{equation}
where~\cite{Mivehvar2021Cavity, Ritsch:2013}
\begin{equation}
    \hat{H}_\mathrm{cav} = -\hbar\Delta\hat a^\dagger\hat a + \hbar (\eta^*\hat{\Theta}_{\rm BH}^\dagger\hat{a} + \eta\hat{a}^\dagger\hat{\Theta}_{\rm BH})\,.
\end{equation}

To eliminate the cavity degrees of freedom, we require that the inequalities $J,u,|\eta|\sqrt{N} \ll |\Delta + \ii \kappa|$ hold. It is instructive to examine first the atom-only master equation of Eq.\ \eqref{eq:eff_master_equation:1} in the adiabatic limit. Thus, we set
\begin{equation}
    \hat{\alpha} \equiv \hat{\alpha}_0 = \frac{\eta}{\Delta + \ii \kappa} \hat{\Theta}_\mathrm{BH}\,,\label{Eq:alpha_0_BH}
\end{equation}
where the expression for $\hat{\alpha}_0$ is derived from Eq.\ \eqref{Eq:ss_alpha_0}. Substituting this expression into Eq.\ \eqref{eq:eff_master_equation:1} yields the adiabatic atom-only master equation $\partial_t\hat\mu =\mathcal L_{0,\rm eff}\hat\mu$, with
\begin{equation}
    \label{L:eff:0}
    \mathcal L_{0,\rm eff}\hat\mu=\frac{1}{{\rm i}\hbar}[\hat H_\mathrm{S} + \hat H_\mathrm{CQED}^\mathrm{(BH)},\hat \mu]+\Gamma\mathcal D[\hat\Theta_{\rm BH}]\hat \mu\,.
\end{equation}
The incoherent term is scaled by the rate $\Gamma = \kappa |\eta|^2/(\Delta^2 + \kappa^2)$. The Hamiltonian $\hat H_{\rm CQED}^\mathrm{(BH)}$ reads explicitly
\begin{equation}
    \label{Eq:H:CQED:1}
    \hat H_{\rm CQED}^\mathrm{(BH)} =\hbar \frac{\Delta|\eta|^2}{\Delta^2 + \kappa^2}\hat{\Theta}_{\mathrm{BH}}^2\,.   
\end{equation}
For $\Delta<0$, the coherent cavity-mediated interaction favors configurations maximizing the expectation value of $\hat\Theta_{\rm BH}^2 = (\sum_j (-1)^j \hat{n}_j)^2$ by either occupying only even or odd sites $j$. 

The atom-only Lindbladian including the diabatic correction $\mathcal L_{1,\rm eff}$ is found by using $\hat \alpha=\hat \alpha_0+\hat \alpha_1$ in Eq.\ \eqref{eq:eff_master_equation:1}, where
\begin{equation}
    \hat{\alpha}_1 = \frac{2J\eta}{(\Delta + \ii \kappa)^2}\sum_{j=1}^{L-1}(-1)^j\left(\hat{b}_j^\dagger\hat{b}_{j+1} - \hat{b}_{j+1}^\dagger\hat{b}_j\right)\label{Eq:alpha_corrected_BH}
\end{equation}
as derived from Eq.\ \eqref{Eq:ss_alpha_1}. This leads to additional terms in the Hamiltonian and the dissipator in comparison to the adiabatic Lindbladian~\eqref{L:eff:0}.

We validate the atom-only master equation by performing the linear decomposition of Lindblad operators, $\mathcal L\hat\rho_\lambda=\lambda\hat\rho_\lambda$, where $\lambda$ are the complex eigenvalues and $\hat\rho_\lambda$ the right eigenvectors~\cite{Englert:2002}. As we detail in Appendix \ref{App:D}, the spectra and the eigenvectors provide full knowledge of the dynamics for a given initial state. It is important to note that the coarse-graining averaging---at the basis of the atom-only master equation---introduces a lower bound on the time resolution, set by the coarse-graining time $\Delta t$. This, in turn, defines an effective energy cutoff $\hbar/\Delta t$, below which the spectrum of $\mathcal L_{\rm cav}$ is expected to be captured by the atom-only description. We thus compare the spectra of eigenvalues of the atom-only Lindbladian $\mathcal L_{0,\rm eff}$ with the eigenvalues of the optomechanical Lindbladian $\mathcal L_{\rm cav}$ in the regime where they are expected to agree. We also determine the spectrum of the atom-only Lindbladian $\mathcal L_{1,\rm eff}$, which includes the first diabatic correction.

We numerically perform the linear decomposition on a small lattice composed of four sites and for two bosons. Figures~\ref{fig:comparison_BH_ED} a) and b) display the eigenvalue spectra for two different values of the atom-cavity coupling strength $\eta\gg J$ but corresponding to the adiabatic parameter $\varepsilon \approx 0.04$ and $\varepsilon\approx 0.36$, respectively. Deep in the adiabatic regime (subplot a)), the three spectra agree within the graphical resolution. As the magnitude of the diabatic corrections increases [see subplot b)], the discrepancy between the atom-only and the optomechanical Lindbladians becomes evident. Nevertheless, the atom-only Lindbladians $\mathcal L_{0,\rm eff}$ and $\mathcal L_{1,\rm eff}$ still capture the slowest-decaying eigenvalues with a relatively small error.

Figures~\ref{fig:comparison_BH_ED} c) and d) display the dynamics of the observable $\langle \hat\Theta_{\rm BH}^2 \rangle$ according to the three Lindblad master equations. This observable is related to the intracavity photon number in the laboratory frame. In the adiabatic regime, subplot a), the diabatic correction remains small at short times, so the predictions according to $\mathcal L_{1,\rm eff}$ do not substantially differ from the adiabatic evolution. At longer times, instead, the adiabatic evolution does not reproduce the dynamics of the full optomechanical simulation. Remarkably, the regime of validity of the atom-only master equation is significantly extended by including the diabatic correction. In subplot d), where the diabatic corrections are substantial, the agreement between the atom-only and the full optomechanical models reduces to shorter times. In the inset, we report the dynamics of the intracavity photon number, $\langle \hat{a}^\dagger\hat{a}\rangle$, obtained from the optomechanical master equation, and the expectation value $\langle \hat{\alpha}^\dagger \hat{\alpha} \rangle$ from the atom-only master equations---equivalent to the photon number in the laboratory frame. This shows that, at very short times, the discrepancy corresponds to the build-up of a finite intracavity field on a timescale shorter than the coarse-graining time.

This comparison shows that, by including systematically higher-order diabatic corrections, one can substantially extend the range of validity of the atom-only model in time in the strong-coupling regime. We remark that the adiabatic approximation will not capture the steady state of the optomechanical model. In particular, the stationary state of the adiabatic, atom-only master equation is an infinite-temperature state that maximizes the entropy. With diabatic corrections, the steady state can be different from the infinite-temperature state, as it is also indicated in Fig.~\ref{fig:comparison_BH_ED} c). That implies that diabatic corrections are essential to describe superradiant stationary states \footnote{Note that for very strong interactions, also diabatic corrections will fail to describe the relaxation into the correct stationary state. This is due to the fact that the basic assumptions of the atom-only master equation are violated since for extremely strong interactions, cavity and atomic degrees of freedom evolve on the same timescale.}. 

We finally emphasize that, despite including the diabatic corrections, the Lindblad form remains preserved. This makes our master equation a powerful framework for describing the dynamics of quantum macroscopic correlations induced by the strong coupling with light.

\section{Conclusion and Outlook}
\label{Sec:Conclusions}

We have presented a theoretical framework that permits to consistently describe the onset of quantum correlations in quantum gases coupled to optical resonators. The framework systematically encompasses cavity-induced phenomena, such as quantum self-organization and cavity cooling of polarizable particles, and can be extended to analyze novel regimes so far inaccessible to existing theories. 

Within this framework, we derived a Lindblad master equation for the mechanical motion of the particles by eliminating the resonator degrees of freedom while retaining slow atom-field correlations. As a result, the master equation captures quantum effects due to multiple photon scattering. The model has a broad range of applications, from cold atoms down to the ultracold regime, from weakly correlated to strongly correlated quantum gases. The description is not limited to the weak-coupling limit between cavity field and atoms, and can cover a relatively wide range of intracavity photon numbers. Since the atoms are here described by their polarizability, this model can also be applied to the dynamics of other polarizable particles, such as molecules in resonators in the regime of coherent scattering, and could be extended to include the rovibrational degrees of freedom in a systematic manner~\cite{Kalaivanan:2021}. 

From the formal point of view, our theory is valid within an ultraviolet cutoff, which determines the shortest timescale $\Delta t$ that the master equation can faithfully describe. This timescale is physical and is determined by the characteristic timescale $\tau_\mathrm{modes}$ of the cavity modes, within which entanglement between the cavity and the atoms is generated~\cite{Halati:2025a}. This establishes a connection with other formal treatments such as the coarse-graining master equation of Refs.\ \cite{Lidar:2001,Majenz:2013}. As for the coarse-graining master equation, we do not need to invoke the rotating-wave approximation to ensure positivity. We further note that the formalism is valid for atoms in multi-mode cavities. It does not have formal limitations on the number of cavity modes and can be extended to a continuum within the coarse-graining approach~\cite{Nathan:2020}.

The theory provides a solid ground for systematically connecting the dynamics of quantum self-organization in cavities with driven-dissipative models of particles interacting via long-range forces~\cite{Defenu:2023}, permitting to identify the experimental parameters that control the dynamics as well as to analyze the dynamics of novel emerging phases of strongly correlated photon-matter systems. Starting from the N-body master equation of cavity QED here derived, a systematic analysis can be performed numerically and analytically, by means of the BBGKY hierarchy extending treatments developed in the semiclassical regime~\cite{Campa:2009,Schuetz:2016,Jaeger:2016} to the full quantum regime. The lowest-order equation is, for ultracold bosons forming a condensate, a generalized Gross-Pitaevskii equation which has been used to predict the stationary and out-of-equilibrium dynamics of quantum gases in two cavity modes, finding very good quantitative agreement~\cite{Baumgaertner:2025}. The corresponding energy functional allows one to identify the free energy landscape of the mean-field solutions, including metastable configurations. This will be subject of future studies. Beyond mean-field, our theory validates analyses based on Bogoliubov excitations of atom-only CQED Hamiltonians~\cite{Sharma:2022} and permits to include in a systematic way the effects of the incoherent dynamics. 

The theoretical model here presented allows one to gain insight into the basic mechanisms at play and establishes the basis for developing protocols to engineer quantum many-body states and dynamics in CQED settings. The formalism can be extended to a variety of platforms, such as optomechanical arrays~\cite{Ludwig:2013,Bemani:2019,Ren:2022} and polaritonic systems~\cite{Restrepo:2014,Kyriienko:2014,Lagoin:2025}, where it may serve as a tool to investigate dynamics beyond mean-field and semiclassical descriptions. The relative simplicity of the model permits to identify the key control parameters, and thereby paves the way towards designing the stationary and out-of-equilibrium dynamics of many-body systems using photon-induced interactions.

\acknowledgments
We are deeply grateful to Corinna Kollath for insightful discussions and for support in preparing the manuscript. The authors also acknowledge discussions with A. Baumg{\"a}rtner, D.\ Dreon, S.\ Hertlein, C.\ M{\'a}ximo, F.\ Piazza, and H.\ Ritsch. 
This work was funded by the Deutsche Forschungsgemeinschaft (DFG, German Research Foundation) – Project-ID 429529648 – TRR 306 QuCoLiMa (“Quantum Cooperativity of Light and Matter", subproject D02) and Project No. 277625399-TRR 185 OSCAR (A4,A5,B4). It also received support from the DFG Forschergruppe WEAVE "Quantum many-body dynamics of matter and light in cavity QED" - Project ID 525057097 and Project number 217124 of the Swiss National Science foundation (SNSF).
This project is funded within the QuantERA II Programme (project "QNet: Quantum transport, metastability, and neuromorphic applications in Quantum Networks"), which has received funding from the EU's Horizon 2020 research and innovation programme under Grant Agreement No. 101017733, as well as from the funding organisations SNSF (Project number 221538) and DFG (Project ID 532771420). 
T.\ D.\ acknowledges support from the Swiss State Secretariat for Education, Research and Innovation (SERI) (project number MB22.00090), and support from the SNSF (project numbers 221538, 217124, 223274). 
C.\ H.\ acknowledges support of the SNSF under Division II grant 200020-219400.  Part of this research was performed at the Kavli Institute for Theoretical Physics (KITP) in Santa Barbara and was supported by grant NSF PHY-2309135 to the Kavli Institute for Theoretical Physics (KITP).
S.B.J. acknowledges support from the DFG under Germany’s Excellence Strategy--Cluster of Excellence Matter and Light for Quantum Computing (ML4Q) EXC 2004/1--390534769.

\section*{Data Availability}
The data and codes that support the findings of this article are openly available at \cite{dataset_zenodo}.

\appendix

\section{Detailed Derivation of the Atom-only Master Equation}
\label{App:A}

In this Appendix, we present a detailed derivation of the atom-only master equation \eqref{eq:eff_master_equation:1}, outlined in Sec.\ \ref{Sec:strong}.

\subsection{Displacement and Useful Relations}

The starting point of the derivation is the displaced master equation introduced in Eq.\ \eqref{eq:displaced_master_equation}, obtained via a transformation using the displacement operator [see Eqs. \ \eqref{eq:D} and \eqref{eq:r_general}]
\begin{equation}
    \hat{D}(t) = \exp[\hat{r}(t)]\,,
\end{equation}
with
\begin{equation}
    \hat{r}(t) = \sum_{n=1}^M (\hat{a}_n^\dagger \hat{\alpha}_n(t) - \hat{\alpha}_n^\dagger(t) \hat{a}_n)\,.
\end{equation}
According to the Baker-Campbell-Hausdorff formula~\cite{Gardiner:QNoise}, a generic operator $\hat{O}$ in the displaced reference frame is given by
\begin{equation}
    \hat{D}^\dagger \hat{O} \hat{D} = \sum_{l=0}^\infty \frac{(-1)^l}{l!}\left[\hat{r},\hat{O}\right]_l\label{eq:DODdagger}\,,
\end{equation}
where $[\hat{r},\hat{O}]_l$ denotes the $l$-order nested commutator, explicitly defined as $[\hat{r},\hat{O}]_0 = \hat{O}$, $[\hat{r},\hat{O}]_1 = [\hat{r},\hat{O}]$, $[\hat{r},\hat{O}]_2 = [\hat{r},[\hat{r},\hat{O}]]$, and so on. To compute the transformed Hamiltonian $\hat{H}_\mathrm{eff}' = \hat{D}^\dagger \hat{H}_\mathrm{eff} \hat{D}  - \ii \hbar \hat{D}^\dagger \partial_t\hat{D}$, the following additional identity is particularly useful:
\begin{equation}
    \hat{D}^\dagger\frac{\partial \hat{D}}{\partial t} = \sum_{l=0}^\infty \frac{(-1)^l}{(l+1)!}\left[\hat{r},\frac{\partial \hat{r}}{\partial t}\right]_l\,.\label{eq:DdaggerderivD}
\end{equation}
This relation follows directly from the operator identity~\cite{Snider:1964}
\begin{equation}
    \frac{\partial}{\partial \lambda} e^{\hat{Y}(\lambda)} = \int_0^1\mathrm{d}x\,e^{(1-x)\hat{Y}}\frac{\partial \hat{Y}}{\partial \lambda}e^{x\hat{Y}}\,,
\end{equation}
for any operator $\hat{Y}(\lambda)$ that is an analytic function of a parameter $\lambda$. In our case, we identify $\hat{Y} \equiv \hat{r}$ and $\lambda \equiv t$.

For the time-dependent atomic operators $\hat{\alpha}_n$ we make the ansatz $\hat{\alpha}_n = \hat{\alpha}_{0,n} + \hat{\alpha}_{1,n}$, where $\hat{\alpha}_{0,n}$ are operators that depend only on the particles' positions, thus, obeying the commutation relations:
\begin{equation}
    [\hat{\alpha}_{0,n},\hat{\alpha}_{0,m}] = [\hat{\alpha}_{0,n},\hat{\alpha}_{0,m}^\dagger] = 0\,,\quad\forall n,m=1,\dots,M\,.\label{App:commutation_rel_alpha_0}
\end{equation}
The operators $\hat{\alpha}_{1,n}$, instead, may depend on both, the particles' positions and momenta.

\subsection{Perturbative expansion}

As motivated in the main text, our objective is to perform a perturbative expansion in the small parameter $\varepsilon$ [see Eq.\ \eqref{Eq:pert_parameter_eps} and the text below]. We identify the scaling
\begin{equation}
    \frac{\hat{H}_\mathrm{S}}{\hbar|\Delta_n + \ii \kappa_n|} \sim \varepsilon^2\,, \quad\frac{\hat{G}_n}{|\Delta_n + \ii \kappa_n|} \sim \varepsilon\,,
\end{equation}
while for the atomic operators $\hat{\alpha}$, we make the assumptions
\begin{equation}
    \hat{\alpha}_{0,n} \sim \varepsilon\,,\quad \hat{\alpha}_{1,n} \sim \varepsilon^{3}\,.
\end{equation}
In view of this perturbative treatment, the transformed Lindbladian $\mathcal{L}_\mathrm{D}$ from Eq.\ \eqref{eq:displaced_master_equation} can be understood as a perturbative series in $\varepsilon$, explicitly expressed as
\begin{equation}
    \mathcal{L}_\mathrm{D} = \sum_{n=0}^\infty\left[\mathcal{L}_\mathrm{D}\right]_{\varepsilon^{n}}\label{app:pert_series_L_D}\,,
\end{equation}
where $\left[\mathcal{L}_\mathrm{D}\right]_{\varepsilon^{n}}$ denotes the collection of all contributions of the displaced Lindbladian of order $\varepsilon^{n}$. Throughout this appendix, we will consistently use this notation to highlight the order of a transformed quantity in $\varepsilon$.

\subsection{Transformed Lindbladian}

In the following, we explicitly compute the different orders of the transformed Lindbladian $\mathcal{L}_\mathrm{D}$.

\subsubsection{Up to order \texorpdfstring{$\varepsilon$}{TEXT}}

We start by computing all terms of the series \eqref{app:pert_series_L_D} up to order $\varepsilon^2$, as indicated by a subscript ``$\leq \varepsilon^2$'':
\begin{equation}
    \left[\mathcal{L}_\mathrm{D}\right]_{\leq \varepsilon^2} \equiv \left[\mathcal{L}_\mathrm{D}\right]_{0} + \left[\mathcal{L}_\mathrm{D}\right]_{\varepsilon} + \left[\mathcal{L}_\mathrm{D}\right]_{\varepsilon^2}\,.
\end{equation}
For this purpose, we can assume $\hat{\alpha}_n = \hat{\alpha}_{0,n}$ by dropping $\hat{\alpha}_{1,n}$ as it would lead to higher-order terms $\sim \varepsilon^{3}$. Using Eq.\ \eqref{eq:DODdagger} and accounting for the commutation relations \eqref{App:commutation_rel_alpha_0}, one can then straightforwardly show that $\hat{D}^\dagger \hat{\Omega}_{nm}\hat{D} = \hat{\Omega}_{nm}$, $\hat{D}^\dagger \hat{G}_{n}\hat{D} = \hat{G}_{n}$, and $\hat{D}^\dagger \hat{a}_{n} \hat{D} = \hat{a}_{n} + \hat{\alpha}_{0,n}$. Furthermore, from Eq.\ \eqref{eq:DdaggerderivD}, we derive the relation
\begin{equation}
    \left[\hat{D}^\dagger\frac{\partial \hat{D}}{\partial t}\right]_{\leq \varepsilon^2} = \sum_{n=1}^M\left(\left(\hat{a}_n^\dagger + \hat{\alpha}_{0,n}^\dagger/2\right)\partial_t\hat{\alpha}_{0,n} - \mathrm{h.c.}\right)\,.
\end{equation}
With these results, the displaced Hamiltonian $\hat{H}_\mathrm{eff}' = \hat{D}^\dagger \hat{H}_\mathrm{eff}\hat{D} - \ii\hbar \hat{D}^\dagger\partial_t\hat{D}$ up to order $\varepsilon^2$ takes the form
\begin{align}
    \left[\hat{H}_\mathrm{eff}'\right]_{\leq \varepsilon} &= \hat{H}_\mathrm{S} + \sum_{n,m=1}^{M}\hbar\hat{a}_n^\dagger\hat{\Omega}_{nm}\hat{a}_{m} + \sum_{n=1}^{M}\hbar(\hat{a}_n^\dagger\hat{\mathcal{E}}_{0,n} + \mathrm{h.c.})\nonumber\\
    &\quad+\sum_{n=1}^{M}\hbar\left[\hat{\alpha}_{0,n}^\dagger\left(\hat{G}_n - \ii \frac{\hbar}{2}\partial_t \hat{\alpha}_{0,n}\right) + \mathrm{h.c.}\right]\label{Eq:H_eff_prime_frozen_particles}\\
    &\quad+ \sum_{n,m=1}^{M}\hbar\hat{\alpha}_{0,n}^\dagger\hat{\Omega}_{nm}\hat{\alpha}_{0,m}\nonumber\,,
\end{align}
where we defined
\begin{equation}
    \hat{\mathcal{E}}_{0,n} = \hat{G}_n + \sum_{m=1}^M \hat{\Omega}_{nm}\hat{\alpha}_{0,m} - \ii \frac{\partial}{\partial t} \hat{\alpha}_{0,n}\,.
\end{equation}
For the displaced dissipator $\mathcal{D}[\hat{D}^\dagger \hat{a}_n \hat{D}]$, we find the expression
\begin{align}
    \left[\mathcal{D}[\hat{D}^\dagger \hat{a}_n \hat{D}]\hat{\tilde{\varrho}}\right]_{\leq \varepsilon^2} &= \mathcal{D}[\hat{a}_n]\hat{\tilde{\varrho}} + \mathcal{D}[\hat{\alpha}_{0,n}]\hat{\tilde{\varrho}}\label{Eq:Dissipator_prime_frozen_particles} + \Bigl(2\hat{a}_n\hat{\tilde{\varrho}}\hat{\alpha}_{0,n}^\dagger\\
    &\quad - \hat{\alpha}_{0,n}^\dagger\hat{a}_n \hat{\tilde{\varrho}} - \hat{a}_n^\dagger\hat{\alpha}_{0,n} \hat{\tilde{\varrho}} + \mathrm{h.c.}\Bigr)\nonumber\,,
\end{align}
with the transformed state $\hat{\tilde{\varrho}} = \hat{D}^\dagger \hat{\varrho} \hat{D}$. With the Hamiltonian \eqref{Eq:H_eff_prime_frozen_particles} and the dissipator \eqref{Eq:Dissipator_prime_frozen_particles}, we have fully determined the displaced Lindbladian up to order $\varepsilon^2$. Our next objective is to determine the operators $\hat{\alpha}_{0,n}$ such that it preserves the factorized form $\hat{\tilde{\varrho}} = \hat{\mu} \otimes \ket{\mathrm{vac}}\bra{\mathrm{vac}}$ of the transformed state. This requirement is satisfied when the following equation holds:
\begin{equation}
    \hat{\mathcal{E}}_{0,n} = \ii \kappa_n\hat{\alpha}_{0,n}\label{app:eom_alpha_0}\,,
\end{equation}
which corresponds to Eq.\ \eqref{eq:easy} in the main text. Under this condition, the transformed Lindbladian to order $\varepsilon^2$ takes the form
\begin{align}
    \left[\mathcal{L}_\mathrm{D}\hat{\tilde{\varrho}}\right]_{\leq \varepsilon^2} &= \frac{1}{\ii \hbar}\Bigl[\hat{H}_\mathrm{eff}^\mathrm{at,0} + \sum_{n,m=1}^{M}\hbar\hat{a}_n^\dagger\hat{\Omega}_{nm}\hat{a}_{m}, \hat{\tilde{\varrho}}\Bigr]\nonumber\\
    &\quad+ \sum_{n=1}^M \kappa_n\mathcal{D}[\hat{a}_n]\hat{\tilde{\varrho}} + \sum_{n=1}^M \kappa_n\mathcal{D}[\hat{\alpha}_{0,n}]\hat{\tilde{\varrho}}\label{Eq:L_D_up_to_eps}\\
    &\quad+ \sum_{n=1}^M 2\kappa_n\left(\hat{a}_n\hat{\tilde{\varrho}}\hat{\alpha}_{0,n}^\dagger - \hat{\alpha}_{0,n}^\dagger\hat{a}_n \hat{\tilde{\varrho}} + \mathrm{h.c.}\right)\nonumber\,,
\end{align}
where $\hat{H}_\mathrm{eff}^\mathrm{at,0}$ is defined in Eq.\ \eqref{eq:H_eff:0}. By construction, the remaining terms acting on the cavity degrees of freedom now vanish when applied to the vacuum state. Therefore, for an initial state of the form $\hat{\tilde{\varrho}} = \hat{\mu} \otimes \ket{\mathrm{vac}}\bra{\mathrm{vac}}$, these terms can be discarded. Furthermore, we can eliminate the cavity degrees of freedom altogether by projecting onto the vacuum state; this leads to the atom-only master equation in Lindblad form presented in Eq.\ \eqref{eq:eff_master_equation:0}, governing the dynamics of the atomic density operator $\hat{\mu} = \bra{\mathrm{vac}}\hat{\tilde{\varrho}}\ket{\mathrm{vac}}$.

\subsubsection{Order \texorpdfstring{$\varepsilon^{3}$}{TEXT}}

The first contribution of the displaced Lindbladian $\mathcal{L}_\mathrm{D}$ originating from the operator $\hat{\alpha}_{1,n}$ appears at order $\varepsilon^{3}$ and is explicitly given by
\begin{equation}
    \begin{aligned}
        &\left[\mathcal{L}_\mathrm{D}\right]_{\varepsilon^{3}}\hat{\tilde{\varrho}}\\
        &= \frac{1}{\ii\hbar}\Bigl[\hbar\sum_{n=1}^M\left(\hat{a}_n^\dagger(\hat{\mathcal{E}}_{1,n} - \ii \kappa_n\hat{\alpha}_{1,n}) + \mathrm{h.c.}\right)\\
        &\quad+ \hbar\sum_{n,m,l = 1}^M\hat{a}_n^\dagger\left([\hat{\alpha}_{1,l}^{\dagger},\hat{\Omega}_{nm}]\hat{a}_l+\hat{a}_l^{\dagger}[\hat{\Omega}_{nm},\hat{\alpha}_{1,l}]\right)\hat{a}_m,\hat{\tilde{\varrho}}\Bigr]\\
        &\quad+ \sum_{n=1}^M 2\kappa_n \Bigl(\hat{a}_n\hat{\tilde{\varrho}}\hat{\alpha}_{1,n}^\dagger - \hat{\alpha}_{1,n}^\dagger\hat{a}_n \hat{\tilde{\varrho}} + \mathrm{h.c.}\Bigr)\label{Eq:L_D_eps_three_half}\,,
    \end{aligned}
\end{equation}
where the operator $\hat{\mathcal{E}}_{1,n}$ is defined as
\begin{equation}
    \hat{\mathcal{E}}_{1,n} = \frac{1}{\hbar}[\hat{H}_\mathrm{S},\hat{\alpha}_{0,n}] + \sum_{m=1}^M \hat{\Omega}_{nm}\hat{\alpha}_{1,m} - \ii \frac{\partial}{\partial t} \hat{\alpha}_{1,n}\,.
\end{equation}
In the expression \eqref{Eq:L_D_eps_three_half}, all terms that couple the vacuum state to higher photon-number states have already been collected in the first line on the RHS. As before, we aim to eliminate this coupling by imposing the condition
\begin{equation}
    \hat{\mathcal{E}}_{1,n} = \ii \kappa_n\hat{\alpha}_{1,n}\label{app:eom_alpha_1}\,,
\end{equation}
which corresponds to the equation of motion for $\hat{\alpha}_{1,n}$ given in Eq.\ \eqref{eq:harder}. Consequently, the $\mathcal{O}(\varepsilon^{3})$ contribution to the displaced Lindbladian vanishes when acting on states of the form $\hat{\tilde{\varrho}} = \hat{\mu} \otimes \ket{\mathrm{vac}}\bra{\mathrm{vac}}$, i.e.,
\begin{equation}
    [\mathcal{L}_\mathrm{D}]_{\varepsilon^{3}}\hat{\tilde{\varrho}} = 0\,.
\end{equation}

\subsubsection{Order \texorpdfstring{$\varepsilon^{4}$}{TEXT}}

The first non-vanishing contribution of $\hat{\alpha}_{1,n}$ to the dynamics of the atomic density operator $\hat{\mu}$ arises at order $\varepsilon^4$, through the term $[\mathcal{L}_\mathrm{D}]_{\varepsilon^4}$. By explicitly deriving this contribution and projecting onto the vacuum state, we obtain
\begin{align}
    &\bra{\mathrm{vac}}([\mathcal{L}_\mathrm{D}]_{\varepsilon^4}\hat{\tilde{\varrho}})\ket{\mathrm{vac}} = \frac{1}{\ii\hbar}\Bigl[\frac{\hbar}{2}\sum_{n=1}^M\left(\hat{G}_n^\dagger\hat{\alpha}_{1,n} + \hat{\alpha}_{1,n}^\dagger\hat{G}_n\right),  \hat{\mu}\Bigr]\nonumber\\
	&+ \sum_{n=1}^M \kappa_n\left(2\hat{\alpha}_{1,n}\hat{\mu}\hat{\alpha}_{0,n}^\dagger - \hat{\alpha}_{0,n}^\dagger\hat{\alpha}_{1,n}\hat{\mu} - \hat{\mu}\hat{\alpha}_{0,n}^\dagger\hat{\alpha}_{1,n} + \mathrm{h.c.}\right)\,.
\end{align}
At first glance, adding this correction to the expression in Eq.\ \eqref{Eq:L_D_up_to_eps} seems to violate the Lindblad form. However, it can be restored by including terms of higher order, specifically of order $\varepsilon^6$. Incorporating such terms then leads to the Lindblad master equation \eqref{eq:eff_master_equation:1}, that is valid up to order $\varepsilon^4$. In a similar way, we may combine both equations of motion for $\hat{\alpha}_{0,n}$ and $\hat{\alpha}_{1,n}$, given in Eqs. \eqref{app:eom_alpha_0} and \eqref{app:eom_alpha_1}, respectively, and include a higher-order term, $(1/\hbar)[\hat{H}_\mathrm{S},\hat{\alpha}_{1,n}]$. This yields a single equation of motion for the composite operator $\hat{\alpha}_n = \hat{\alpha}_{0,n} + \hat{\alpha}_{1,n}$.

\section{Cavity cooling}
\label{App:Cavity cooling}

In this Appendix, we provide further details on the case study of cavity cooling, discussed in Sec. \ref{Sec:weak}.

\subsection{Optomechanical master equation}

We consider a single atom that interacts with a single-mode cavity and is driven by a pump orthogonal to the cavity axis. In the dispersive regime, $|\Delta_\mathrm{a}| \gg \gamma$, the optomechanical master equation, Eq.\ \eqref{eq:opto_master_equation}, simplifies to
\begin{equation}
    \label{eq:full_master_equation_cavity_cooling}
    \frac{\partial}{\partial t}\hat{\varrho} = \frac{1}{\ii\hbar}[\hat{H}_\mathrm{eff},\hat{\varrho}] + \mathcal{L}_\kappa\hat{\varrho}\,,
\end{equation}
where the dissipator for a single cavity mode reads $\mathcal{L}_\kappa = \kappa(2\hat{a}\hat{\varrho}\hat{a}^\dagger - [\hat{a}^\dagger\hat{a},\hat{\varrho}]_+)$. Without direct drive of the cavity mode, $\hat{H}_\mathrm{pump}^{(c)} = 0$, the Hamiltonian is of the form $\hat{H}_\mathrm{eff} = \hat{H}_\mathrm{S} + \hat{H}_\mathrm{mode} + \hat{V}_\mathrm{eff}$. The free Hamiltonian of the cavity mode reads explicitly $\hat{H}_\mathrm{mode} = - \hbar \Delta \hat{a}^\dagger\hat{a}$. We assume that the particle is tightly confined in the transverse direction, thus, restricting their motion to the cavity axis, say the $x$-axis. The Hamiltonian of the particle then reads $\hat{H}_\mathrm{S} = \hat{p}^2/(2m)$. By virtue of this transversal confinement, the laser, which propagates perpendicular to the motional axis, is uniform and the mechanical effects of light are solely due to cavity-atom interactions: $\hat{V}_\mathrm{eff} = \hbar U \, \hat{\Theta}_{11}\hat{a}^\dagger\hat{a} + \hbar (\eta \, \hat{a}^\dagger\hat{\Theta}_{1\mathrm{p}} + \mathrm{h.c.})$, with $\hat{\Theta}_{1\mathrm{p}} = \cos(k\hat{x})$, $\hat{\Theta}_{11} = \cos^2(k\hat{x})$, and $k$ is the wave number of the cavity mode.

\subsection{Atom-only master equation}

To derive the atom-only master equation, we first express the atomic operator
\begin{equation}
    \label{eq:alpha_expansion_momentum}
    \hat{\alpha} = \sum_{p,p'}\alpha_{p,p'}\ket{p}\bra{p'}
\end{equation}
in the eigenbasis of the mechanical Hamiltonian $\hat{H}_\mathrm{S}$, that is, the momentum eigenstates $\ket{p}$, satisfying $\hat{H}_\mathrm{S}\ket{p} = p^2/(2m)\ket{p}$. The coefficients $\alpha_{p,p'}$ obey the coupled differential equations [see Eq.\ \eqref{eq:alpha:gen}]
\begin{equation}
    \begin{aligned}
        \ii\frac{\partial}{\partial t}\alpha_{p,p'} =& \left(\frac{p^2 - (p')^2}{2m\hbar} - \Delta + \frac{U}{2} - \ii \kappa\right)\alpha_{p,p'}\\
        & + \frac{U}{4}(\alpha_{p-2\hbar k,p'} + \alpha_{p+2\hbar k,p'})\\
        & +\frac{\eta}{2}(\delta_{p',p-\hbar k} + \delta_{p',p+\hbar k})\,.
    \end{aligned}
\end{equation}
In the weak-coupling limit, $|\eta| \ll \kappa$, this equation can be solved for its steady state. Assuming further $|U| \ll |\eta|$, we can discard the contributions coming from the dispersive shift, such that the steady state takes the form
\begin{equation}
    \alpha_{p,p'} = \alpha_-(p')\delta_{p, p' - \hbar k} + \alpha_+(p')\delta_{p, p' + \hbar k}\,,
\end{equation}
with $\alpha_\pm$ given in Eq.\ \eqref{alpha:p}. Plugging the expression for the coefficients in the expansion \eqref{eq:alpha_expansion_momentum}, we obtain the expression presented in the main text in Eq.\ \eqref{alpha_op:p}. The atom-only master equation, Eq.\ \eqref{Eq:mu}, is here given by 
\begin{equation}
    \frac{\partial}{\partial t}\hat{\mu} = \frac{1}{\ii \hbar}[\hat{H}_\mathrm{eff}^\mathrm{at},\hat{\mu}] + \kappa\mathcal{D}[\hat{\alpha}]\hat{\mu}\,,    
\end{equation}
with
\begin{equation}
    \hat{H}_\mathrm{eff}^\mathrm{at} = \hat{H}_\mathrm{S} + \frac{\hbar}{2}\left(\eta\,\hat{\alpha}^\dagger\hat{\Theta}_{1\mathrm{p}} + \mathrm{h.c.}\right)\,.
\end{equation}

\subsection{Rate equations}
\label{App:Cavity cooling:Occupation dynamics}

In order to get insight into the cooling dynamics, we analyze the dynamical evolution of the momentum state occupations $\Pi_p = \bra{p}\hat{\mu}\ket{p}$. Their equation of motion is found by projecting the atom-only master equation onto the momentum state $\ket{p}$:
\begin{equation}
    \label{Eq:pop_dyn_eom}
    \frac{\partial}{\partial t}\Pi_p = \frac{1}{\ii\hbar}\bra{p}[\hat{H}_\mathrm{eff}^\mathrm{at},\hat{\mu}]\ket{p} + \kappa\bra{p}\left(\mathcal{D}[\hat{\alpha}]\hat{\mu}\right)\ket{p}\,.
\end{equation}
Using that $\hat{\Theta}_{1 \mathrm{p}}\ket{p} = (\ket{p - \hbar k} + \ket{p + \hbar k})/2$, we obtain
\begin{equation}
    \label{Eq:pop_dyn_coherent_part}
    \begin{aligned}
        &\bra{p}[\hat{H}_\mathrm{eff}^\mathrm{at},\hat{\mu}]\ket{p}\\
        =&~\frac{\hbar}{4} \Bigl([\eta\alpha_-^*(p) + \eta^*\alpha_+(p -2\hbar k)]\bra{p - 2 \hbar k}\hat{\mu}\ket{p}\\
        &+ [\eta\alpha_+^*(p) + \eta^*\alpha_-(p + 2\hbar k)]\bra{p + 2 \hbar k}\hat{\mu}\ket{p} - \mathrm{c.c.}\Bigr)\,,
    \end{aligned}
\end{equation}
where $\mathrm{c.c.}$ denotes the complex conjugate, and
\begin{widetext}
    \begin{equation}
        \label{Eq:pop_dyn_incoherent_part}
        \begin{aligned}
            \bra{p}\left(\mathcal{D}[\hat{\alpha}]\hat{\mu}\right)\ket{p} =& 2|\alpha_+(p - \hbar k)|^2\Pi_{p - \hbar k} + 2|\alpha_-(p + \hbar k)|^2\Pi_{p + \hbar k}- 2(|\alpha_-(p)|^2 + |\alpha_+(p)|^2)\Pi_p\\
            & + 2\Bigl(\alpha_+(p-\hbar k)\alpha_-^*(p+\hbar k)\bra{p-\hbar k}\hat{\mu}\ket{p+\hbar k} + \mathrm{c.c.}\Bigr)\\
            & - \Bigl(\alpha_-^*(p)\alpha_+(p-2\hbar k)\bra{p-2\hbar k}\hat{\mu}\ket{p} + \alpha_+^*(p)\alpha_-(p+2\hbar k)\bra{p+2\hbar k}\hat{\mu}\ket{p} + \mathrm{c.c.}\Bigr)\,.
        \end{aligned}
    \end{equation}
\end{widetext}
Generally, the populations $\Pi_p$ couple to coherences $\bra{p'}\hat{\mu}\ket{p}$, $p' \neq p$, as visible from Eqs.\ \eqref{Eq:pop_dyn_coherent_part} and \eqref{Eq:pop_dyn_incoherent_part}. In the weak-coupling limit, with a cavity loss rate that fulfills $|\eta|/\kappa \ll 1$, this coupling can be discarded and we can describe the dynamics solely in terms of occupations. For the dynamics of the latter, we can thus neglect the coherent part, Eq.\ \eqref{Eq:pop_dyn_coherent_part}, and restrict to the incoherent part, Eq.\ \eqref{Eq:pop_dyn_incoherent_part}, simplifying to
\begin{align}
    &\bra{p}\left(\mathcal{D}[\hat{\alpha}]\hat{\mu}\right)\ket{p}\nonumber\\
    &\approx 2|\alpha_+(p - \hbar k)|^2\Pi_{p - \hbar k} + 2|\alpha_-(p + \hbar k)|^2\Pi_{p + \hbar k}\nonumber\\
    &- 2(|\alpha_-(p)|^2 + |\alpha_+(p)|^2)\Pi_p\,.
\end{align}
With this, Eq.\ \eqref{Eq:pop_dyn_eom} becomes the equation of motion \eqref{eq:P} given in the main text, with the rates $r_\pm(p) = 2\kappa|\alpha_\pm(p)|^2$.

\subsection{Steady-state distribution}
\label{App:Cavity cooling:Steady-state distribution}

We now analyse the steady-state probability $\Pi_p$ for large momenta $p$. We first observe that $r_+(-p)=r_-(p)$; therefore, the steady-state distribution is symmetric about $p=0$, and we can focus on the populations with $p\ge 0$. By imposing detailed balance, the stationary distribution satisfies the relation
\begin{align}
    \Pi_{p+\hbar k}=\Pi_{p}\frac{r_+(p)}{r_-(p+\hbar k)}.
\end{align}
Recursively applying this relation leads to the equation connecting the population at momentum $p$ with the population at momentum $p'=p+n\hbar k$, with $n > 0$:
\begin{align}
    \Pi_{p'}=\Pi_p\prod_{s=1}^{n}\frac{r_+(p+(s-1)\hbar k)}{r_-(p+s\hbar k)}\,.
\end{align}
For $s\gg 1$, one can approximate the ratio with a Taylor expansion truncated in first order in $1/s$:
\begin{align}
\frac{r_+(p+(s-1)\hbar k)}{r_-(p+s\hbar k)}\approx 1+2\frac{\Delta}{s \, \omega_\mathrm{R}}\,.
\end{align}
In the following, we simplify 
\begin{align}
    \ln(\Pi_{p'})\sim\sum_{s=1}^n\ln\left(\frac{r_+(p+(s-1)\hbar k)}{r_-(p+s\hbar k)}\right)
\end{align}
for very large $p'$ and by performing a Taylor expansion of the summands for very large $s$ and $n$. Here, $\ln$ denotes the natural logarithm. Taking the infrared cutoff $s_0$ in the sum, we obtain $\sum_{s=s_0}^{n}1/s\sim \ln(n)$ for very large $n$ and
\begin{align}
    \ln(\Pi_{p'})\sim2\frac{\Delta}{\omega_\mathrm{R}}\ln(n)\,.
\end{align}
For large momenta $p'=p + n\hbar k\approx n\hbar k$, we then obtain the behavior of the populations at the tails of the distribution:
\begin{align}
    \Pi_{p'}\sim  n^{2\frac{\Delta}{\omega_\mathrm{R}}}= C\left(\frac{p^{\prime 2}}{2m\hbar\omega_\mathrm{R}}\right)^{\Delta/\omega_\mathrm{R}}\,, 
\end{align}
with $C$ a constant; see also Refs.~\cite{Niedenzu:2011,Griesser:2012}.
Note that this result implies that the probability density is not normalizable if $\Delta\geq-\omega_\mathrm{R}/2$. In fact, in this regime, the coupling with the cavity 
heats the atoms. Moreover, the mean kinetic energy is infinite for $\Delta\geq-3\omega_\mathrm{R}/2$ due to the contributions of the power-law tails to the integral.  

\section{Bose-Hubbard Model with Cavity-mediated Interactions}
\label{App:C}

In this Section, we provide further details on the Bose-Hubbard model with cavity-mediated long-range interactions, studied in Sec. \ref{subsec:strong_coupling_Bose-Hubbard}.

\subsection{Second-quantized Hamiltonian}

Consider a quantum gas of identical atoms, which is conveniently described in the second quantization formalism. We denote by $\hat\psi(\vec{r})$ the atomic field operator obeying $[\hat\psi(\vec{r}),\hat{\psi}^\dagger(\vec{r}')]_\pm=\delta(\vec{r}-\vec{r}')$ and $[\hat\psi(\vec{r}),\hat{\psi}(\vec{r}')]_\pm=0$, with $[~]_\pm$ indicating the anticommutator ($+$) or commutator ($-$) depending on whether the atoms are fermions or bosons, respectively. Let the total Hamiltonian be of the form
\begin{equation}
    \hat H=\hat H_{\rm ext} + \hat{H}_{\rm coll} + \hat H_{\rm cav}\,,
\end{equation}
with the external Hamiltonian
\begin{equation}
    \label{Eq:H_ext_second_quantization}
  \hat{H}_\mathrm{ext}=\int {\rm d}^3r~\hat\psi^\dagger(\vec{r})\left(-\frac{\hbar^2}{2m}\vec{\nabla}^2 + W(\vec{r})\right)\hat \psi(\vec{r})\,,
\end{equation}
and $\hat{H}_{\rm coll}$ the s-wave scattering term for bosons:
\begin{equation}
    \label{Eq:H:coll_second_quantization}
    \hat{H}_{\rm coll}= \hbar\frac{u'}{2}\int {\rm d}^3r~\hat{\psi}^\dagger(\vec{r})\hat{\psi}^\dagger(\vec{r})\hat{\psi}(\vec{r})\hat{\psi}(\vec{r})\,,  
\end{equation}
with $u' > 0$ the scaling amplitude. Within this formalism, the second-quantized form of the operators $\hat{\Theta}$ [see Eqs.\ \eqref{eq:def_Theta_operator}], entering in the Hamiltonian $\hat{H}_{\rm cav}$ [see Eq.\ \eqref{eq:Hcav}] originating from the coupling to cavity modes, reads
\begin{equation}
    \hat{\Theta}_{nm} = \int {\rm d}^3r~\hat\psi^\dagger(\vec{r})f_n^*(\vec{r})f_m(\vec{r})\hat\psi(\vec{r})\,,
\end{equation}
with $n,m\in\{\mathrm{p},1,2,\dots,M\}$.

\subsection{Dissipative Bose-Hubbard model of CQED}

In the following, we provide a systematic derivation of the atom-only, extended Bose-Hubbard model. For simplicity, we consider a single cavity mode. The atoms' motion is assumed to be tightly bound to the cavity axis, say the $x$-axis, through an external potential, creating effectively a one-dimensional system; they are further pumped by a laser field orthogonal to the motional axis. Additionally, a one-dimensional optical lattice $W(\hat{x})=V_0\cos^2(k_0\hat{x})$ confines the atoms along the cavity axis with lattice depth $V_0$ and wave number $k_0$, such that the periodicity is $a=\pi/k_0$. In the single-band approximation, the one-dimensional atomic field operator can be written as
\begin{equation}
    \hat{\psi}(x)=\sum_i w_i(x) \hat{b}_i\,,
\end{equation}
where $w_i(x)$ is the (real) Wannier function centered at the lattice site $x_i=ia$ and $\hat{b}_i$ annihilates a boson at site $i$, such that $[\hat{b}_j,\hat{b}_i^\dagger] = \delta_{i,j}$ and $[\hat{b}_j,\hat{b}_i] = 0$. Using the Wannier decomposition, the external Hamiltonian \eqref{Eq:H_ext_second_quantization} gives rise to a local energy and a hopping term, while the contact interactions, Eq.\ \eqref{Eq:H:coll_second_quantization}, turn into an onsite repulsion term. The resulting Bose-Hubbard Hamiltonian is given in Eq.\ \eqref{Eq:H:BH} with coefficients $J=-\int {\rm d}x~w_i(x)(-(\hbar^2/2m)\partial_x^2+W(x))w_{i+1}(x)$ and $u=u'\int{\rm d}x~(w_i(x))^4$, see Ref.\ \cite{Jaksch:1998}. In the one-dimensional case, we can approximate $f_1^*(\vec{r})f_\mathrm{p}(\vec{r}) \approx \cos(kx)$, where $k$ is the cavity mode wave number, and the operator $\hat\Theta_{1\mathrm{p}}$ takes the form
\begin{equation}
    \label{Eq:Theta:BH}
    \hat{\Theta}_\mathrm{BH}=\sum_j \left(Z_j \hat{n}_j + Y_j\left(\hat{b}_j^\dagger \hat{b}_{j+1} + \hat{b}_{j+1}^\dagger \hat{b}_{j}\right)\right)\,.
\end{equation}
Here, $Z_j=\int {\rm d}x~w_j(x)\cos(kx)w_j(x)$, $Y_j=\int {\rm d}x~w_j(x)\cos(kx)w_{j+1}(x)$, and $\hat{n}_j = \hat{b}_j^\dagger \hat{b}_j$. In writing Eq.\ \eqref{Eq:Theta:BH} we have dropped the terms beyond nearest-neighbors, consistently with the expansion of the Bose-Hubbard term \eqref{Eq:H:BH} and in line with the tight-binding approximation.

Using Eq.\ \eqref{Eq:Theta:BH} into Eq.\ \eqref{Eq:H:CQED:1}, we obtain a CQED Hamiltonian describing global density-density interactions, density-mediated tunneling, and global bond-bond interactions~\cite{Fernandez:2010,Chanda:2022}:
\begin{equation}
    \label{Eq:CQED:BH}
    \begin{aligned}
          \hat{H}_{\rm CQED}^\mathrm{(BH)} &= \frac{\hbar|\eta|^2\Delta}{\Delta^2 + \kappa^2}\sum_{i,j}\Bigl(Z_iZ_j\hat{n}_i \hat{n}_j+Z_iY_j\hat n_i \hat B_j\\
          &~+Y_iZ_j\hat{B}_i\hat{n}_j  + Y_iY_j\hat{B}_i\hat{B}_j\Bigr)\,,  
    \end{aligned}
\end{equation}
where $\hat B_j=\hat{b}_j^\dagger \hat{b}_{j+1} + \hat{b}_{j+1}^\dagger \hat{b}_{j}$. The dissipative part in Eq.\ \eqref{L:eff:0} acquires an interesting form, showing that cavity decay, being nonlocal, establishes correlations between densities ($\hat{n}_j$), as well as between bonds ($\hat{B}_j$), and between densities and bonds of every pair of sites:
\begin{equation}
    \begin{aligned}
        \mathcal D[\hat{\Theta}_\mathrm{BH}]\hat\mu &= \sum_{i,j}Z_iZ_j\left(2\hat n_i\hat \mu\hat n_j
        -[\hat n_j\hat n_i,\hat \mu]_+\right)\label{Eq:ZiZj}\\
        &\quad+\sum_{i,j}Y_iY_j\left(2\hat B_i\hat \mu\hat B_j
        -[\hat B_j\hat B_i,\hat \mu]_+\right)\\
        &\quad+\sum_{i,j}Y_iZ_j\left(2\hat B_i\hat \mu\hat n_j
        -[\hat n_j\hat B_i,\hat \mu]_++{\rm h.c.}\right)\,.
    \end{aligned}
\end{equation}
The specific form of the coefficients $Z_i,Y_i$ depends on the ratio between the periodicity of the optical lattice and of the cavity field, determined by $k/k_0$. Incommensurate ratios can give rise to exotic glass phases, see Refs.\ \cite{Habibian:2013,Habibian:2013b,Niederle:2016} for the Hamiltonian case. Mobility edges, with signatures of many-body localization, have been reported in Ref.\  \cite{Sierant:2019,Kubala:2021}. 

The dynamics in the presence of the incoherent term, instead, is largely unexplored. Within the operator approach, the incoherent term gives rise to the input noise term in the Heisenberg-Langevin equation \cite{Larson:2008a,Habibian:2013a}. In most treatments of many-body CQED, it has been neglected, assuming $\kappa\ll |\Delta|$. 

\section{Spectral Decomposition of the Lindblad Master Equation}
\label{App:D}

We shortly recall the spectral decomposition of the master equation, see, e.g., Ref.\ \cite{Englert:2002}. For a generic Lindblad superoperator $\mathcal L$ governing the dynamics of a density operator $\hat{\rho}$, the CPTP (completely positive trace preserving) map $\Lambda$ connecting the evolution $\hat{\rho}(t)$ with the initial state $\hat{\rho}(0)$, $\hat{\rho}(t)=\Lambda[\hat{\rho}(0)]$, can be written as $\Lambda[\hat{\rho}(0)]=\exp(\mathcal{L} t)\hat{\rho}(0)$. This can be cast in the form
\begin{equation}
    \label{eq:spectrum}
    \hat{\rho}(t)=\sum_\lambda c_\lambda {\rm e}^{\lambda t}\hat\varrho_\lambda\,,
\end{equation}
where $\lambda$ are the complex eigenvalues of $\mathcal{L}$ such that $\mathcal{L}\hat{\varrho}_\lambda = \lambda\hat{\varrho}_\lambda$, with $\hat{\varrho}_\lambda$ right eigenstate, while $c_\lambda={\rm Tr}\{\check{\varrho}_\lambda^\dagger\hat{\rho}(0)\}$ is the projection onto the left eigenstate $\check{\varrho}_\lambda$. The latter fulfill the eigenvalue equation $\check{\varrho}_\lambda \mathcal{L}= \lambda\check{\varrho}_\lambda$ and satisfy the orthonormality condition ${\rm Tr}\{\check{\varrho}_{\lambda'}^\dagger\hat{\varrho}_\lambda\}=\delta_{\lambda',\lambda}$. The spectral decomposition of Eq.\ \eqref{eq:spectrum} assumes completeness, which is often satisfied but not {\it a priori} warranted.

Note that, in the limit of vanishing hopping amplitude $J=0$, one can compute analytically the eigenstates and eigenvalues of the full Lindbladian $\mathcal L_{\rm cav}$ of the extended Bose-Hubbard model of Eq.\ \eqref{Eq:L:full}, as the local densities are conserved quantities~\cite{Halati:2020b, Halati:2025a}. 

\bibliography{bibliography}

\end{document}